\documentclass[floatfix,superscriptaddress,preprint]{revtex4-1}
\usepackage[english]{babel}
\usepackage{graphicx}
\usepackage{float}
\usepackage{dcolumn}
\usepackage{bm}

\usepackage{gensymb}
\usepackage{epsf}
\newcommand\twoFigureSize{0.47\linewidth}
\newcommand\twoImageSize{0.495\linewidth}
	
\newcommand\SCALE{0.62}

\begin{document}
	
	\title { Effect of Heterogeneity in Models of El-Ni\~{n}o Southern Oscillations }
	
\author{Chandrakala Meena\footnote{email : chandrakala@iisermohali.ac.in}} 
\author{Shweta Kumari\footnote{email :  eshwetak@iisermohali.ac.in}}
\affiliation{Indian Institute of Science Education and Research (IISER) Mohali, Knowledge City, SAS Nagar, Sector $81$, Manauli PO $140$ $306$, Punjab, India}
\author{ Akansha Sharma\footnote{email : akanshas@buffalo.edu}}
\affiliation{University at Buffalo,Buffalo New York 14261, USA}
\author{Sudeshna Sinha\footnote{email : sudeshna@iisermohali.ac.in}}
\affiliation{Indian Institute of Science Education and Research (IISER) Mohali, Knowledge City, SAS Nagar, Sector $81$, Manauli PO $140$ $306$, Punjab, India}


	\begin{abstract}
		The emergence of oscillations in models of the El-Ni\~{n}o effect is of utmost relevance.
		Here we investigate a coupled nonlinear delay differential system modeling the El Ni\~{n}o/ Southern Oscillation (ENSO) phenomenon,
		which arises through the strong coupling of the ocean-atmosphere system. In particular, we study the temporal patterns of the sea
		surface temperature anomaly of the two sub-regions. For identical sub-regions we typically observe a co-existence of amplitude and
		oscillator death behavior for low delays, and heterogeneous oscillations for high delays, when inter-region coupling is weak.
		For moderate inter-region coupling strengths one obtains homogeneous oscillations for sufficiently large delays and amplitude death
		for small delays. When the inter-region coupling strength is large, oscillations are suppressed altogether, implying that strongly
		coupled sub-regions do not exhibit ENSO-like oscillations. Further we observe that larger strengths of self-delay coupling favours oscillations,
		while oscillations die out when the delayed coupling is weak. This indicates again that delayed feedback, incorporating oceanic wave
		transit effects, is the principal cause of oscillatory behaviour. So the effect of trapped ocean waves propagating in a basin with
		closed boundaries is crucial for the emergence of ENSO. Further, we show how non-uniformity in delays, and difference in the strengths
		of the self-delay coupling of the sub-regions, affect the rise of oscillations. The trends are similar to the uniform system. Namely,
		larger delays and self-delay coupling strengths lead to oscillations, while strong inter-region coupling kills oscillatory behaviour.
		The difference between the uniform case and the non-uniform system, is that amplitude death and homogeneous oscillations are
		predominant in the former, while oscillator death and heterogeneous oscillations are commonly found in the latter. Interestingly,
		we also find that when one sub-region has low delay and another has high delay, under weak coupling the oscillatory sub-region
		induces oscillations in sub-region that would have gone to a steady state if uncoupled. Thus we find that coupling sub-regions has
		a very significant effect on the emergence of oscillations, and strong coupling typically suppresses oscillations, while weak coupling
		of non-identical sub-regions can induce oscillations, thereby favouring ENSO.


	\end{abstract}
	
	\maketitle
	\section{Introduction}
	
	 The atmospheric phenomena known as the  El-Ni\~{n}o event, occurring at intervals of two to seven years, has garnered widespread popular interest due to its global impact ranging from environment to economics. The term  El-Ni\~{n}o typically signifies a very large scale warm event, and this dramatic change in sea surface temperature (SST) is one phase of the  El-Ni\~{n}o Southern Oscillations (ENSO), that is an irregular cycle of coupled ocean temperature and atmospheric pressure oscillations across the equatorial Pacific region. Since 1899 twenty nine El-Ni\~{n}o events have been recorded, namely the cycle of hot and cold phases has an average periodicity of approximately 3.7 years.

	In normal years, SST of the western Pacific Ocean is high and pressure is low compared to the eastern Pacific Ocean. Due to high SST of western region, evaporation increases and moist air rises. As a result condensation happens and high rainfall occurs in western region. Because of cold SST in the east and high level pressure, less rainfall occurs. A pressure gradient in the east and west pacific ocean induces circulations of trade winds. With the effect of circulating trade winds the depth of thermocline gradient changes. In normal conditions, thermocline is deeper in the western and shallower in the eastern Pacific region. In the early stage of El-Ni\~{n}o the circulation of trade winds gets weaker. When El-Ni\~{n}o becomes very strong the circulation of trade winds flips its direction. As a result the thermocline depth becomes almost the same in both east and west Pacific Ocean. In contrast to El-Ni\~{n}o, La-Ni\~{n}a is the extreme phase of the normal condition. Namely, El-Ni\~{n}o is a warm phase  and La-Ni\~{n}o a cold phase of ENSO. 

	Over the past several decades extensive studies have attempted to understand and predict the mechanism and behavior of ENSO \cite{Andrews,Encyclopedia,Holton_Hakim,Soon1,Soon2, basicmodel,review}. In addition to detailed models involving large-scale simulations, there have also been significant attempts to gain understanding of the underlying mechanisms of ENSO through low order models (LOM), describing the phenomenon qualitatively \cite{delay,AmJPhys,Compare_model,Lin,Frauen}. Here the very complex situation is reduced to a system of ordinary differential equations involving a few variables of interest, through a series of approximations and hypotheses. Usually, the LOM so obtained describes the average dynamics of the phenomenon. These models typically assume a positive ocean-atmosphere feedback in the equatorial eastern and central Pacific, leading SST to the warm state responsible for El-Ni\~{n}o. Such models, though simple, are important, as they offer interpretations of the oscillatory nature of ENSO \cite{review,basicmodel}. Important examples of this class of minimal, yet effective, models are the {\em recharge oscillator} and {\em delayed oscillator}.

	The recharge oscillator model (ROM) focuses on the heat content of the tropical Pacific ocean. Prior to  El-Ni\~{n}o, warm water of the western Pacific ocean flushed towards eastern Pacific ocean and upper ocean heat content or warm water volume over the eastern Pacific ocean tend to build up (or charge ) gradually, and during El-Ni\~{n}o warm water is flushed toward (or discharged to) western Pacific and then warm water slowly builds up again (recharge) before occurrence of next El-Ni\~{n}o in the eastern Pacific. This recharge and discharge process of heat content leads to a transition phase in which the entire equatorial Pacific thermocline depth changes. During the discharge process thermocline depth is anomalously shallow in the east region and this allows cold waters to be pumped into the surface layer by climatological up-welling, leading to the cold phase (La-Ni\~{n}o). It is the recharge-discharge process that makes the coupled ocean-atmosphere system oscillate on interannual time scales. Mathematically in ROM, the Eastern Pacific sea surface temperature (SST) and mean equatorial Pacific thermocline depth, which are key variables in ENSO, are represented by a damped oscillator, with the thermocline depth and SST playing the roles of position and momentum in the system. This framework can be readily applied to comprehend the basic El-Ni\~{n}o/La-Ni\~{n}a cycle and is consistent with the high potential predictability of El Ni\~{n}o.

	Additionally, such damped oscillations can be considered to be driven by atmospheric noise, such as the westerly wind bursts in the tropical western Pacific \cite{bianucci, recharge2}. Significantly, such stochastic extensions appear to explain the irregularity of El-Ni\~{n}o, and suggests that it is virtually unpredictable at long time scales. Specifically such models have shown that multiplicative noise may destabilize the ENSO oscillator, alter the mean evolution of ENSO, amplify the ensemble spread and make it initial-condition dependent \cite{recharge2}.

	The second important deterministic low order model, the {\em delayed action oscillator model} \cite{delay}, is the focus of this work. Delayed negative feedback models provide a very good, yet simple, representation of the basic mechanism of ENSO-like oscillations.  An important feature of this class of models is the inclusion of a delayed feedback which incorporates oceanic wave transit effects, namely the effect of trapped ocean waves propagating in a basin with closed boundaries. 
	
	Specifically, the delayed-action oscillator model has three terms, and is a first  order nonlinear delay differential equation for the temperature anomaly $T$, i.e. the deviation from a suitably long term average temperature, given by:

	\begin{equation}
		\frac{dT}{dt}=k T- b T^3- A T(t-\Delta)
		\label{basic}
	\end{equation}
	Here the coupling constants are $k$, $b$ and $A$, with $\Delta$ being the delay. The first term represents a positive feedback in the ocean-atmosphere system, working through advective processes giving rise to temperature perturbations that result in atmospheric heating. The heating in turn leads to surface winds driving the ocean currents which then enhance the anomalous values of $T$. The second term is a damping term, due to advective and moist processes, that limits the temperatures from growing without bound. The delay term arises from considerations of equatorially trapped ocean waves propagating across the Pacific and interacting back after a time delay, determined by the width of the Pacific basin and wave velocities. The strength of this interaction, relative to the nondelayed feedback is given by $A$.

	We will consider the dimensionless form of this equation \cite{AmJPhys}:
 	\begin{equation}
		\frac{dT}{dt}=T-T^3-\alpha T(t-\delta)
		\label{main1}
	\end{equation}
	
	where time in Eqn.~1 has been scaled by $k$, temperature by $\sqrt{b/k}$. The dimensionless constants $\alpha = A/k$ and $\delta = k \Delta$ \cite{AmJPhys}. This  model allows multiple steady states and when these fixed points become unstable, self-sustained oscillations emerge. Thus this class of models provide a simple explanation of ENSO, and provides insights on the key features that allow the emergence of oscillatory behavior. 

	\section{Coupled Delayed-Oscillator Model}
	The delayed-oscillator model given by Eqns~1-2 above consider a single region with strong atmospheric-ocean coupling, namely some typical representative region in the Pacific Ocean. Now, we will consider scenarios in which other regions of the Pacific are incorporated in the model. Specifically, we will explore models mimicking the coupling of regions along the equator, where one expects varying self-delay coupling strengths in the sub-regions, as well as varying (possibly strong) delay times \cite{AmJPhys}. We first describe our coupled model in detail here, and then in the following sections investigate the emergence of oscillations in this class of models. Our main motivation will be to explore the effect of coupling and sub-region heterogeneity on the ENSO, an aspect not explored adequately in earlier studies.
	
	Consider two coupled sub-regions, given by following dimensionless delay differential equations, as introduced in \cite{AmJPhys}:	
	\begin{eqnarray} 
     \label{main}
	\frac{dT_{1}}{dt}= T_{1}-T_{1}^3-\alpha_{1} T_{1}(t-\delta_{1})+\gamma T_{2}\\ \nonumber
	\frac{dT_{2}}{dt}= T_{2}-T_{2}^3-\alpha_{2} T_{2}(t-\delta_{2})+\gamma T_{1}
	\end{eqnarray} 
	
	Here $T_{i}$, $\delta_{i}$ and $\alpha_{i}$ with $i=1,2$ are the scaled temperature 
	anomaly, self-delay, strength of the self-delay of each sub-region, and $\gamma$ is
	the inter-region coupling strength between the two regions. The form of the coupling
	term models the situation where if one region is cooler than the other, then the flow
	of energy across their common boundary will result in heating one sub-region and
	cooling the other. We explore the phenomenon arising in this system, and focus in
	particular on the effect of non-uniformity on the emergent oscillations.
	In this work we will first consider the case of identical sub-regions, i.e.
	$\alpha_{1}=\alpha_2$ and $\delta_{1}=\delta_2$. This implies that the two regions 
	are geographically close-by, and the distance from the western boundary is 
	approximately same, with the same losses and reflection properties for both regions. 
	Since the distance from the western boundary is similar, the transient time taken by 
	the oceanic waves is also expected to be similar, and so the time period of ENSO-like
	oscillations is the same in the sub-regions. After exploring the case of coupled identical sub-regions
	in depth, we will go on to examine the case of non-identical sub-regions, i.e. 
	$\alpha_{1} \ne \alpha_2$ and $\delta_{1} \ne \delta_2$. Here the distance from the 
	western boundary is different for the sub-regions and therefore the transient times 
	taken by the oceanic waves are different in the sub-regions. So the time period of 
	oscillations in each sub-region now  is expected to depend crucially on the values of 
	the parameters $\alpha_{i}$, $\delta_{i}$ and $\gamma$, and may or may not be the same
	. In the sections below, we give details of the rich variety of temporal patterns 
	arising in this coupled system.
		
	\section{Dynamical Patterns in Coupled identical sub-regions}
\begin{figure}[H]
			\centering 
			\includegraphics[scale=\SCALE]{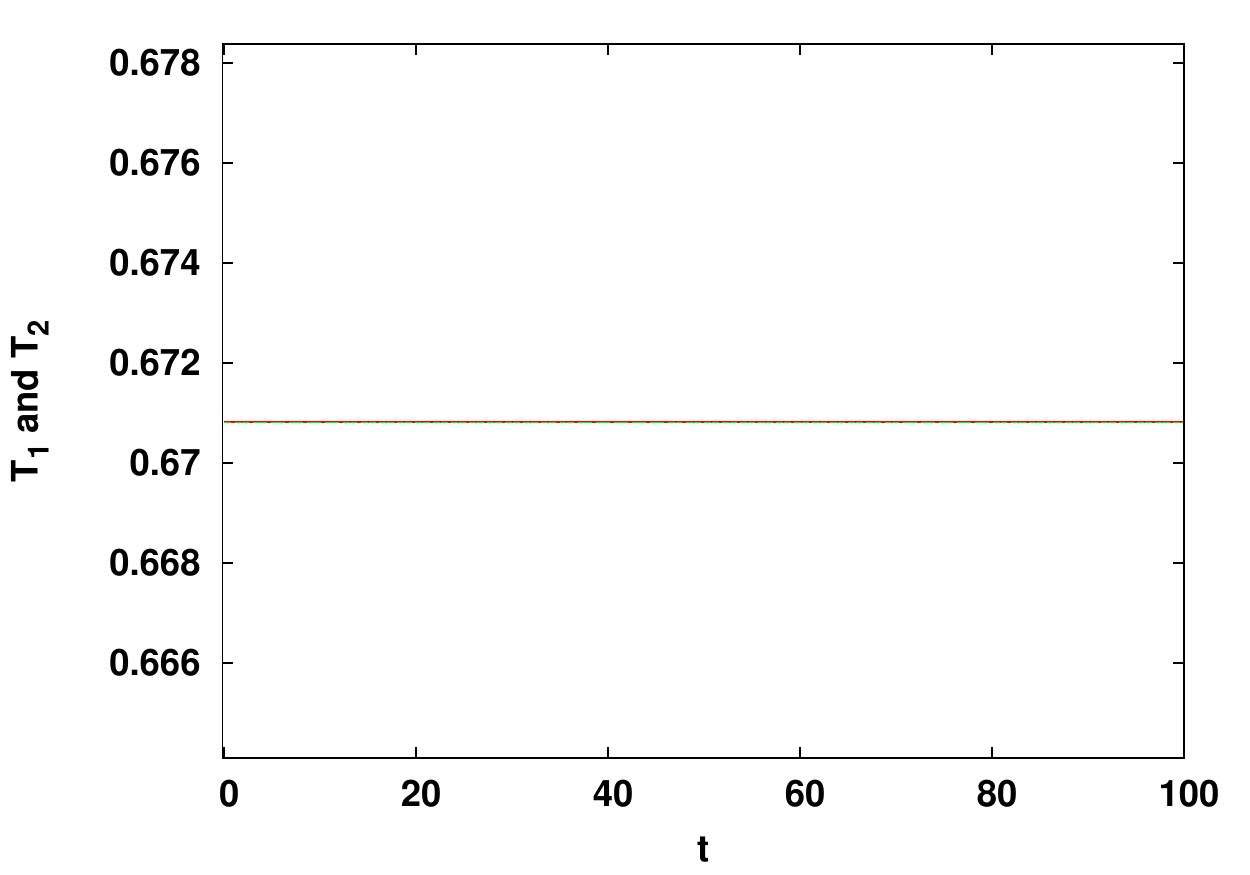}
			\includegraphics[scale=\SCALE]{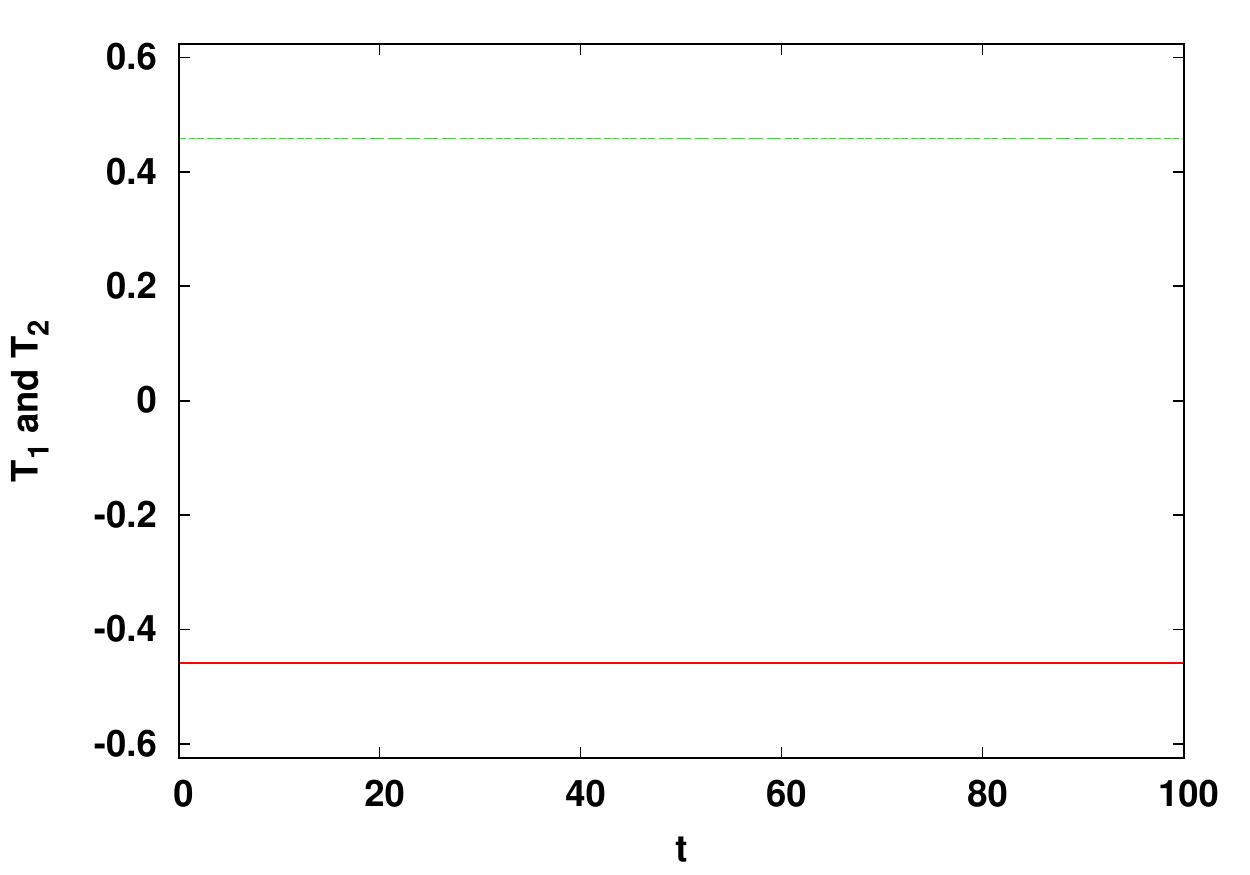}
			\caption{Temporal evolution of the temperature anomalies of the two sub-regions $T_1$ (in red) and $T_2$ (in green) with $\alpha_{1} = \alpha_{2} = 0.75$, $\delta_{1}=\delta_{2}=1$, and inter-region coupling (right) $\gamma=0.2$  and (left) $\gamma=0.05$, showing amplitude death and oscillator death behavior respectively.}
			\label{identical1}
	\end{figure}

\begin{figure}[H]
			\centering 
			\includegraphics[scale=\SCALE]{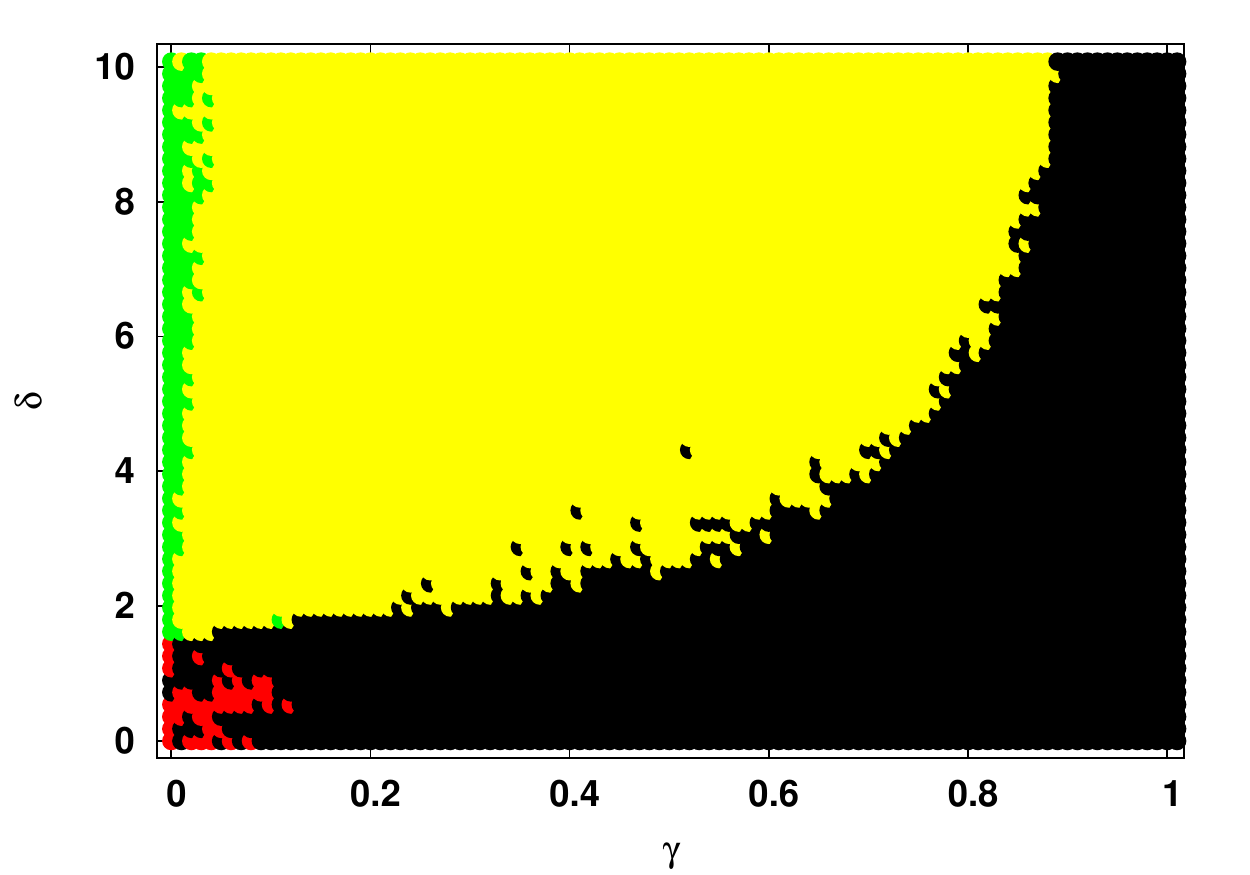}
			\caption{Phase diagram showing the dynamics of the temperature anomaly in mean 
			sea surface temperature of a sub-region ($T_1/T_2$), arising from a typical 
			initial state, with respect to inter-region coupling $\gamma$ and delay 
			$\delta_{1}=\delta_{2}=\delta$. Here the strength of delayed coupling in the 
			two regions is $\alpha_{1} = \alpha_{2} = 0.75$, in Eqn. \ref{main}. The black 
			color represents amplitude death, red represents oscillator death, yellow 
			represents homogeneous oscillations and green represents heterogeneous 
			oscillations.}
			\label{identical3}
	\end{figure}

	First we consider the dynamics of two identical sub-regions, with uniform delays and coupling strengths. This holds true if the two adjacent sub-regions have similar reflection properties and similar losses, and the distance to the western boundary is approximately the same for the two sub-regions as they are close-by geographically. 
	We investigate the range of dynamical behaviour that emerges from a typical initial condition as delays and self-delay coupling strengths, and the inter-region coupling strengths, are varied (cf. Fig.\ref{identical3}). We observe four distinct types of behaviour:
	
	(i) Amplitude Death (AD) : here both regions go to a single steady state  \cite{AD_OD}. See left panel of Fig. \ref{identical1}  for a representative example.
	
	(ii) Oscillation Death (OD): here the sub-regions go to different steady states \cite{AD_OD}. See right panel of Fig. \ref{identical1}  for a representative example.
	
\begin{figure}[H]
			\centering 
			\includegraphics[scale=\SCALE]{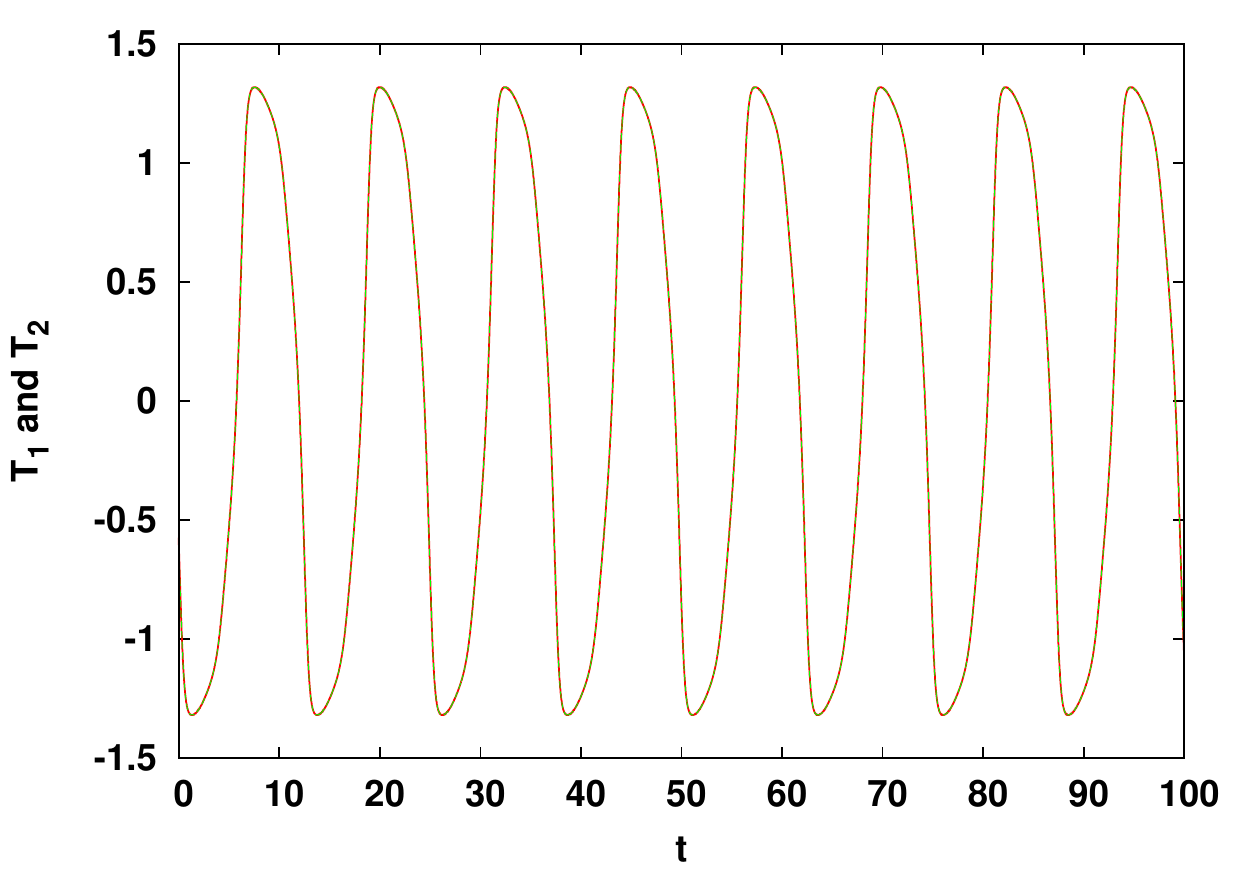}
			\includegraphics[scale=\SCALE]{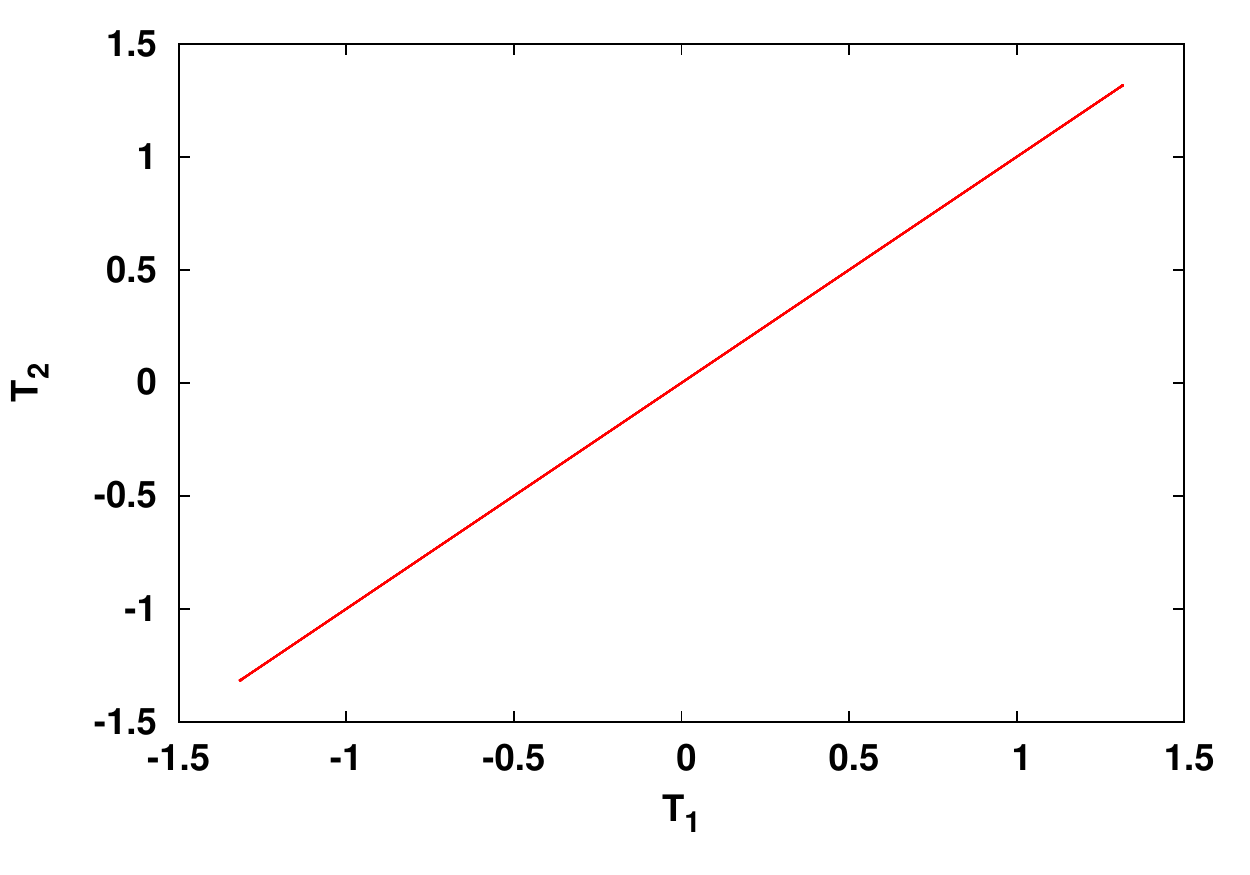}
			\caption{Temporal evolution of the temperature anomalies of the two 
			sub-regions $T_1$ (in red) and $T_2$ (in green) in the left panel, and the 
			corresponding phase portrait in the $T_1-T_2$ plane in the right panel, for 
			$\alpha_{1} = \alpha_{2} = 0.75$, $\delta_{1}=\delta_{2}=4$ and $\gamma=0.1$ 
			in Eqn. \ref{main}.}
	       \label{identical2}		
	\end{figure}

	(iii) Homogeneous oscillations : here the regions oscillate synchronously and there is no phase or amplitude difference between the oscillations. See Fig. \ref{identical2} for a representative example.
	
	(iv) Heterogeneous oscillations : here the oscillatory patterns are complex, and the oscillations in the two sub-regions differ in either phase or amplitude, or both. Further, the oscillations may be irregular for certain parameters. See Figs. \ref{delta_alpha_1}, \ref{delta_alpha_2}, \ref{patterns}  for representative examples.
	
\begin{figure}[H]
			\centering 	
			\includegraphics[scale=\SCALE]{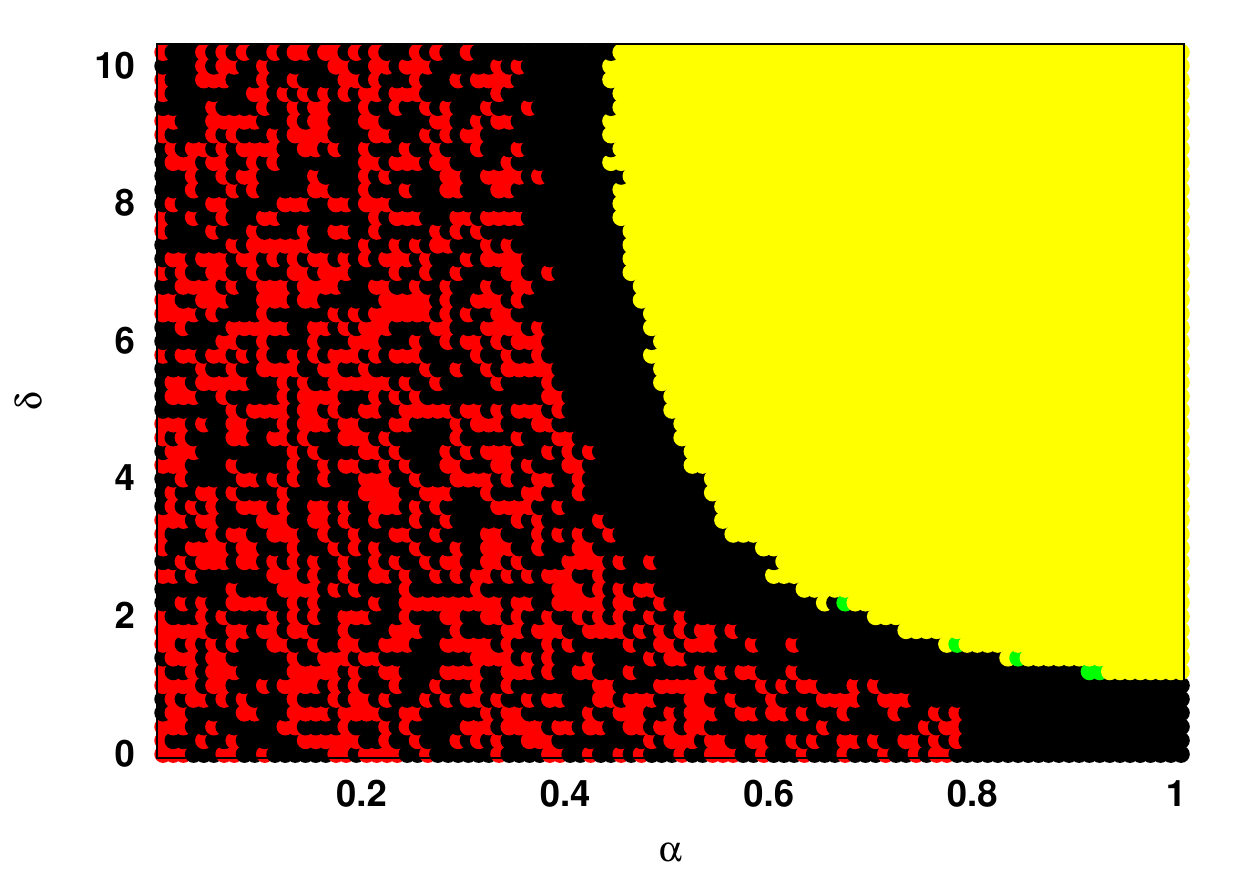}
			\includegraphics[scale=\SCALE]{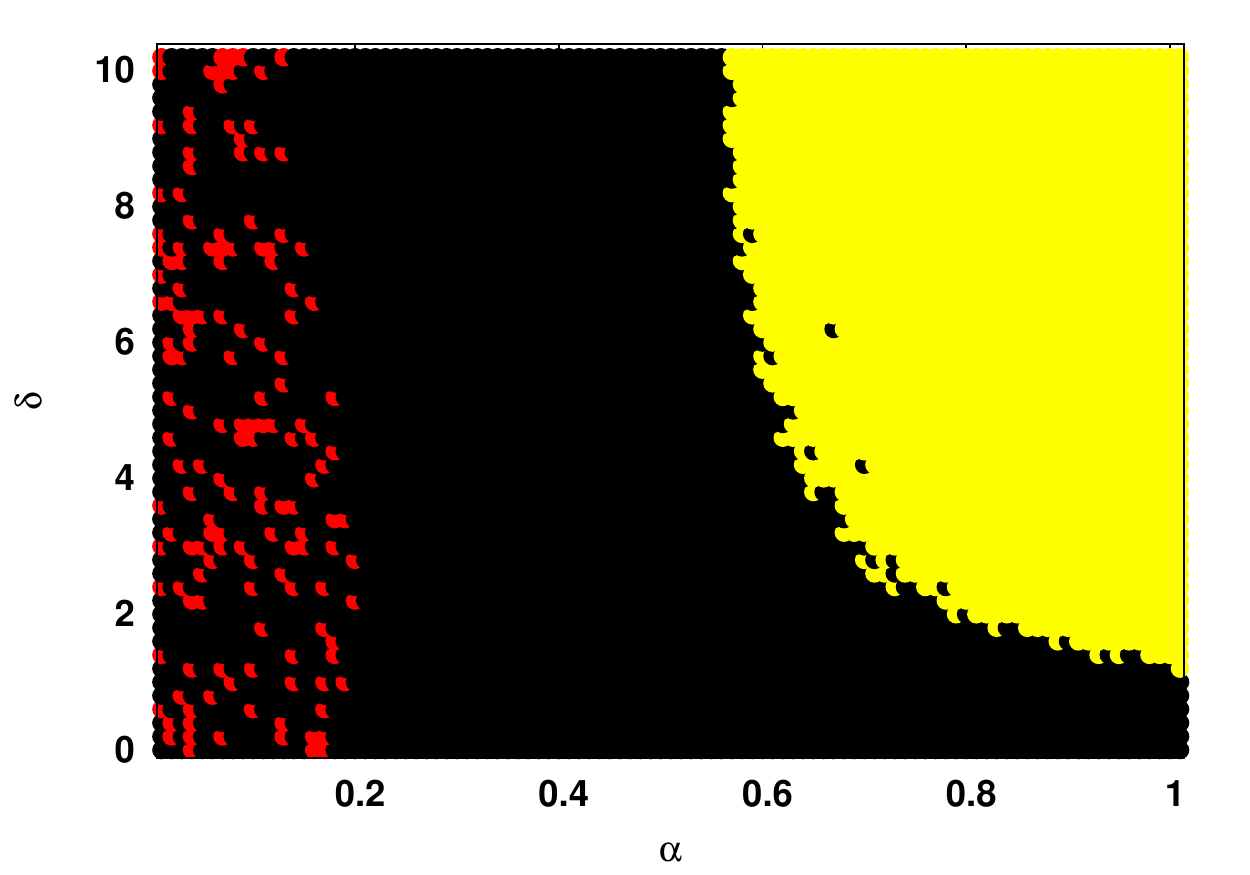}
			\caption{Phase diagram like in Fig. \ref{identical3} for
			$\gamma=0.1$ (left) and $\gamma=0.4$ (right), in Eqn. \ref{main}.}	\label{identical4}		
	\end{figure}

	It is evident from the representative cases in Figs. \ref{identical3} and \ref{identical4} that oscillations emerge as the delay $\delta$ and strength of self-delay coupling $\alpha$ increases, and as inter-region coupling strength $\gamma$ decreases. Importantly, as compared to a single region model, oscillations arise for larger values of delay in the two coupled sub-regions model. This implies that coupling of sub-regions yields smaller parameter regions giving rise to El-Ni\~{n}o oscillations.
	
	\begin{figure}[H]
			\centering 
			\includegraphics[scale=\SCALE]{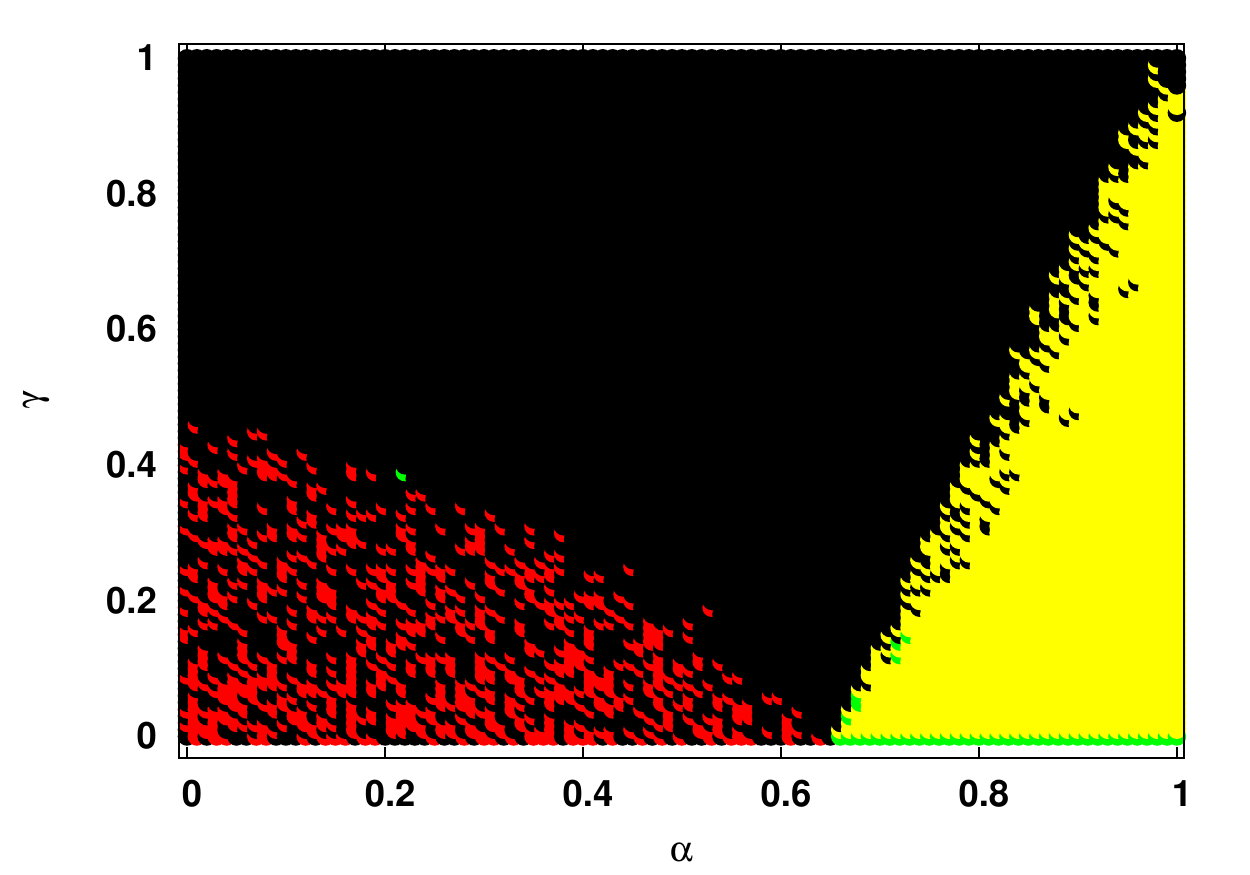}
			\includegraphics[scale=\SCALE]{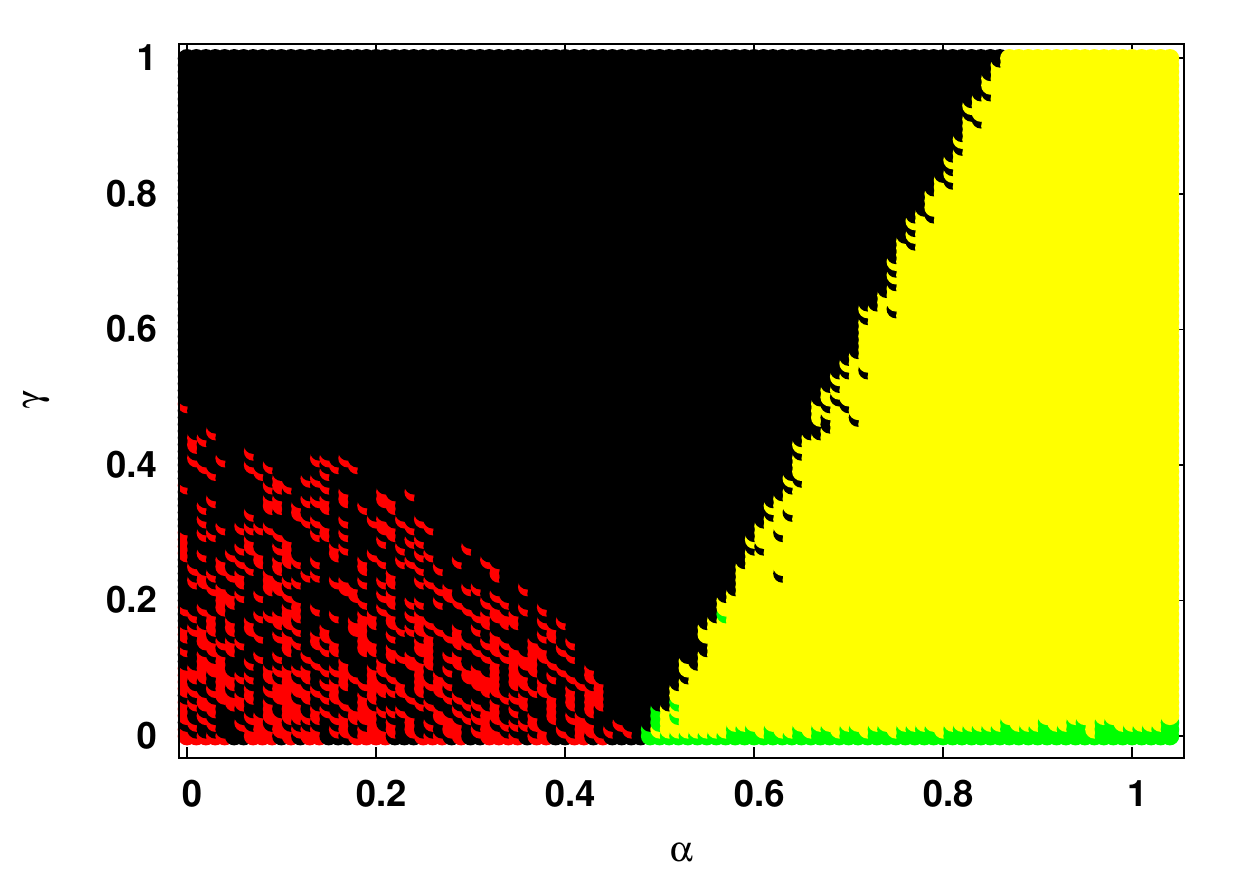}
			\caption{Phase diagram like in above Figs. \ref{identical3}- \ref{identical4}
			  for the delay $\delta_{1}=\delta_{2}=\delta$ equal 
			  to $2$ (left) and $4$ (right), in Eqn. \ref{main}. 
			  } \label{identical5}				
	\end{figure}
	
	Note that in our model system we have observed that the number of attractors and their basins of attraction depend upon the values of  parameters. For instance, when $\alpha_{1}=\alpha_{2}=0.75, \delta=1$, we find four steady states for $\gamma=0.1$ and two steady states for $\gamma>=0.2$. The value of the fixed points depend on the values of the inter-region coupling strength $\gamma$. When $\alpha_{1}=\alpha_{2}=0.5, \delta=1$ and $\gamma=0.1$, we observe four steady states. When initial values of both regions are positive or both are negative, then the system in both regions approach the same steady state. However, when the initial states are different, namely one region is positive and the other negative, then they approach different steady states, i.e. one positive and one negative steady state. For the typical case of $\alpha_1 \ne \alpha_2$, each region has two fixed points and two oscillator states, with the attractors being different in the two regions. Generically, in such cases there is a complex co-existence of attractors.

	Specifically for instance, in Figs. \ref{identical3}-\ref{identical5}, the parameter
	regions with inter-mixed colors  implies co-existing dynamical states, such as
	co-existing amplitude death (AD) and oscillator death (OD) states where there are red
	dots interspersed in the black region. As the strength of the inter-region coupling
	$\gamma$ increases, co-existence of AD and OD decreases. Further, the region of 
	amplitude death increases (cf. Fig. \ref{identical4}), implying that the ENSO-like 
	oscillations are less likely when two sub-regions are strongly coupled. 
	We also observe that as delay $\delta$ increases, co-existence of AD and OD decreases,
	 and the parameter region supporting oscillatory behaviour increases (cf. Fig. 
	\ref{identical5}). For instance, when $\delta=2$ oscillations emerge for self-delay
	coupling strength $\alpha \geq 0.65$, while for $\delta=4$ oscillations emerge in the
    systems with $\alpha \geq 0.48$. So longer delays, namely longer oceanic wave transit
    times, favour ENSO-like oscillations.
				   
	\section{Analysis}
	On the assumption that delay $\delta$ is small ($\delta < 1$), we can consider that 
	the delayed temperature anomaly $T (t-\delta)$ to be approximated by $T (t) - \delta \frac{d T_{t}}{dt}$. Hence we need to solve the following dynamical equations:
	\begin{eqnarray}
		( 1 - \alpha \delta) \frac{dT_{1}}{dt} &=& T_{1} - T_{1}^3 - \alpha  T_{1} +\gamma T_{2}\\
		( 1 - \alpha \delta) \frac{dT_{2}}{dt} &=& T_{2} - T_{2}^3 - \alpha  T_{2}  +\gamma T_{1} 
		\label{small_delay}
	\end{eqnarray}
		
\begin{figure}[H]
			\centering 
			\includegraphics[scale=\SCALE]{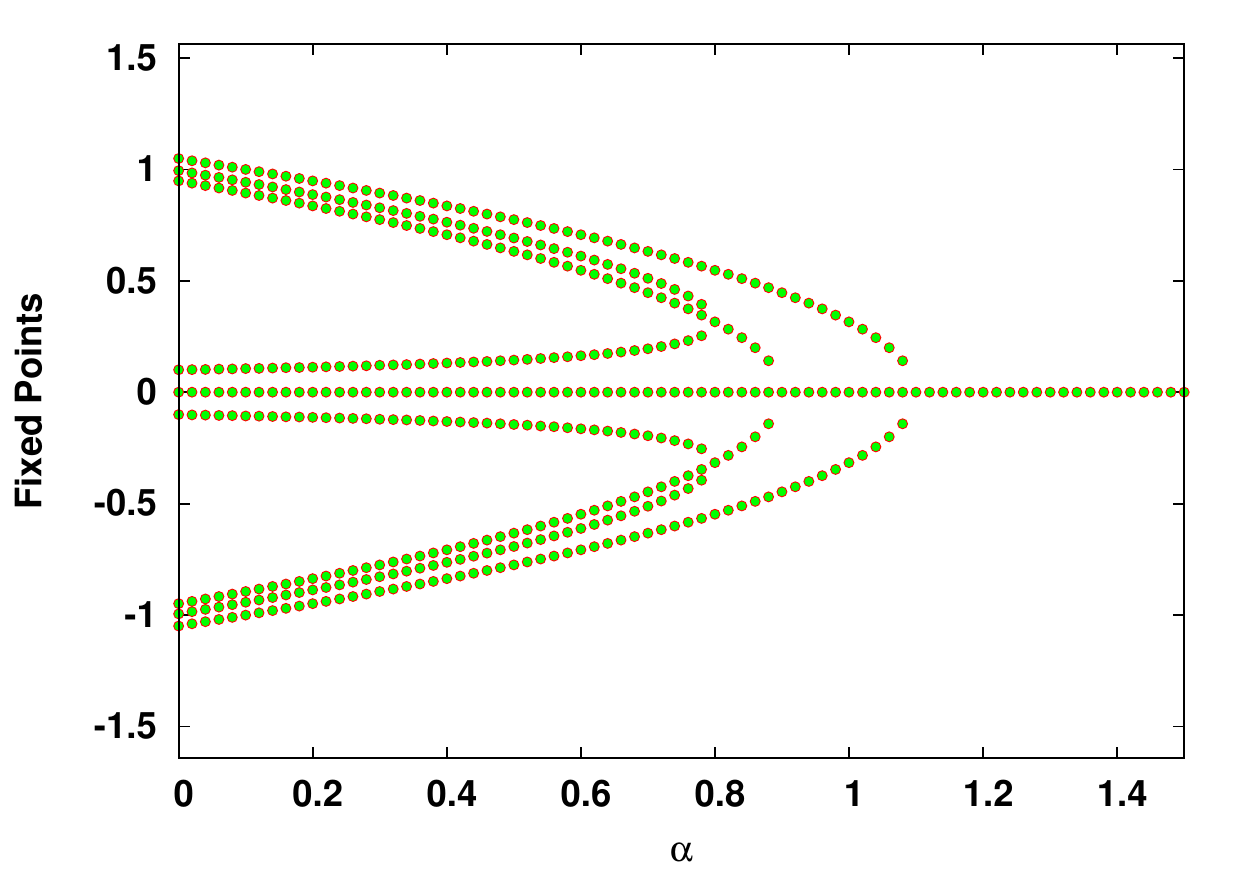}
			\includegraphics[scale=\SCALE]{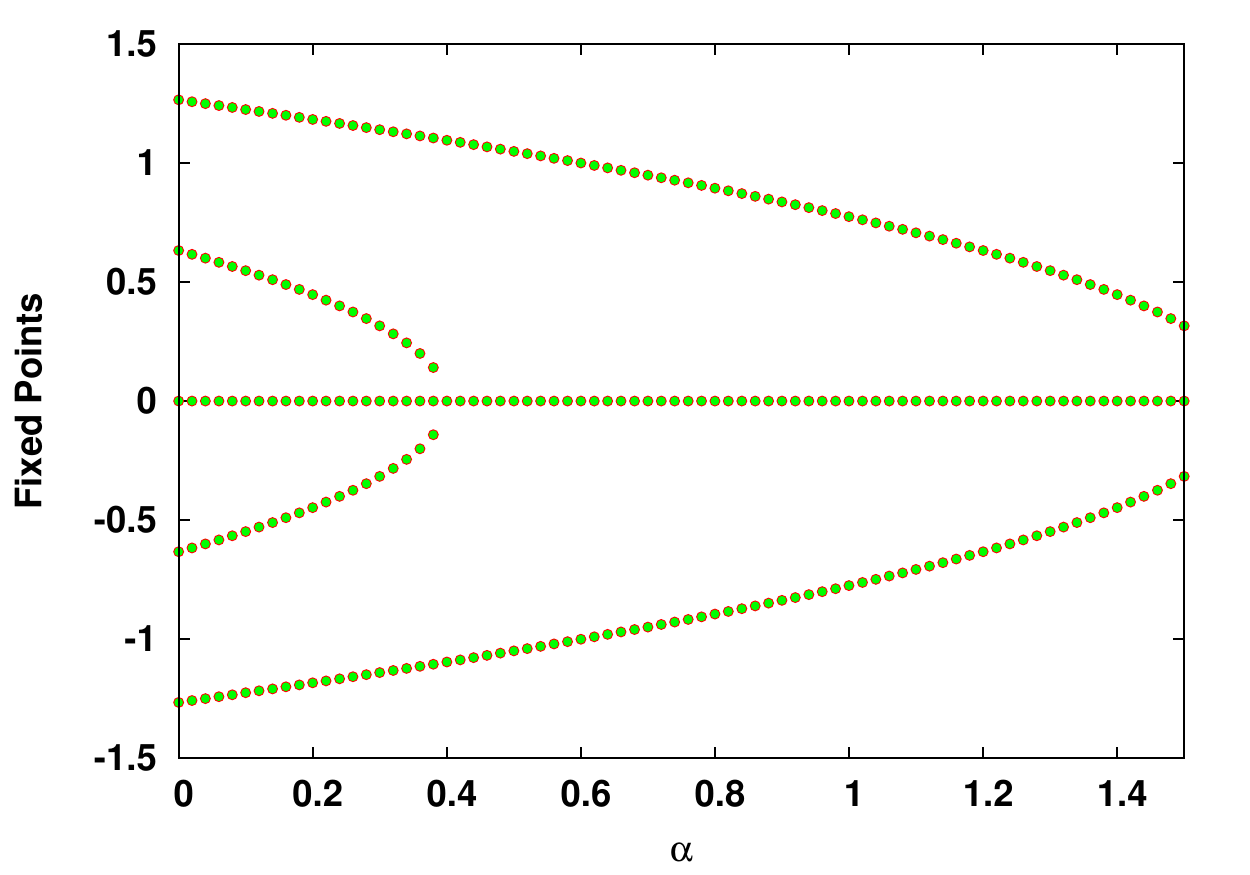}
			\caption{Fixed point solutions arising from Eqn. \ref{small_delay} versus strength of the self-delay coupling $\alpha$ for inter-region coupling $\gamma$ equal to $0.1$ (left) and $0.6$ (right).}
			\label{stability_analysis}		
		\end{figure}

		The Jacobian of the system above is given by:
		\[ J = \frac{1}{1-\alpha\delta}\left( \begin{array}{cc}
		1-3T_{1}^2-\alpha & \gamma  \\
		\gamma & 1-3T_{2}^2-\alpha \end{array} \right).\] 
		The linear stability of the different fixed points that arise in this system, under varying parameters, are determined by the eigen values of $J$. 
		Fig. \ref{stability_analysis} shows the number of steady states for representative parameters $\alpha$ and $\gamma$. A noticeable trend is that as the inter-region coupling $\gamma$ increases, one obtains fewer fixed points at the same value of self-delayed coupling strength $\alpha$. For instance, for small $\alpha$, nine fixed points exist for $\gamma=0.1$, while only five fixed points are there for $\gamma=0.6$. The other feature is that the number of fixed point solutions decreases with $\alpha$, e.g. for $\gamma = 0.1$, there are nine fixed points for small $\alpha$ and only one for large $\alpha$.
	
		For $\gamma=0.1$, the stability of the fixed points for different $\alpha$ is as follows: (a) for $0 \le \alpha<0.8$ there are $4$ stable nodes, $4$ saddle points and $1$ unstable node, (b) for $0.8 \le \alpha < 0.9$, there are $2$ stable nodes, $2$ saddle points and $1$ unstable node, (c) for $0.9 \le \alpha < 1.1$ there are $2$ stable nodes and $1$ saddle point, and lastly (d) for $\alpha \ge 1.1$ we obtain only one fixed point which is a stable node. For $\gamma=0.6$ one obtains a similar stability pattern, shifted down the $\alpha$ axis, starting with $5$ fixed point solutions. So the nature of the fixed points obtained from analyzing Eqn. \ref{small_delay} explains our observation of amplitude death and oscillator death in different regions of parameter space.
	
		\section{Effect of non-uniform self-delay coupling strengths}
\begin{figure}[H]
			\centering 
			\includegraphics[scale=\SCALE]{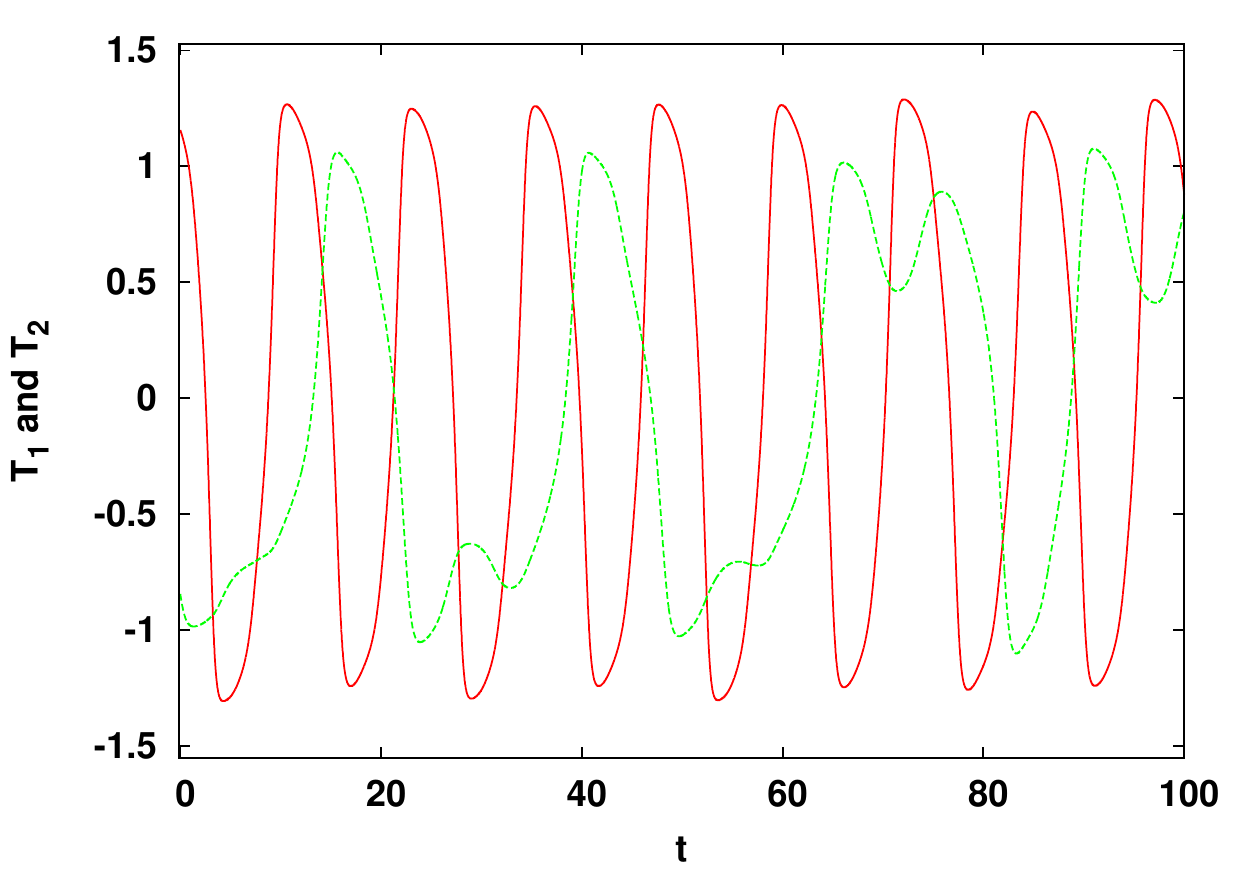}
			\includegraphics[scale=\SCALE]{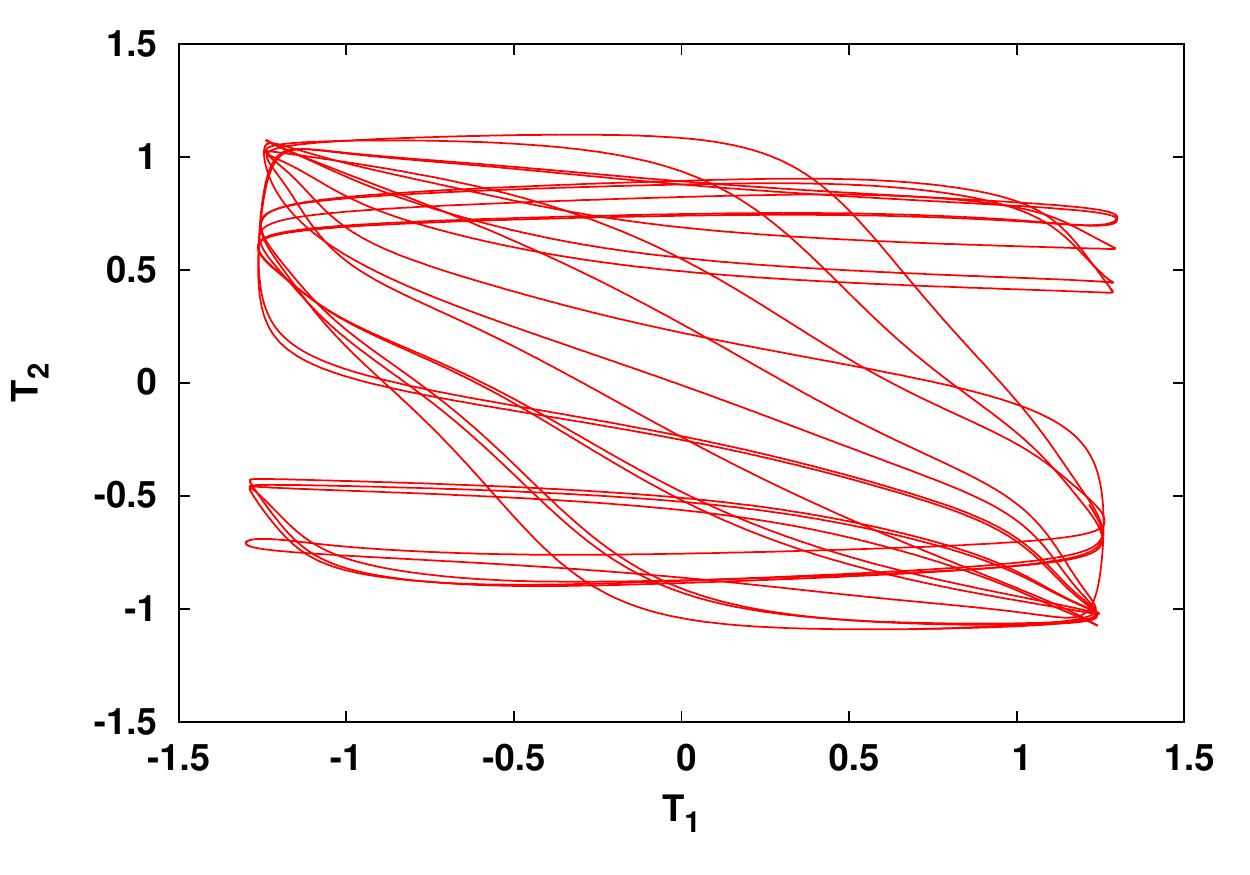}
			(a) 
		
			\includegraphics[scale=\SCALE]{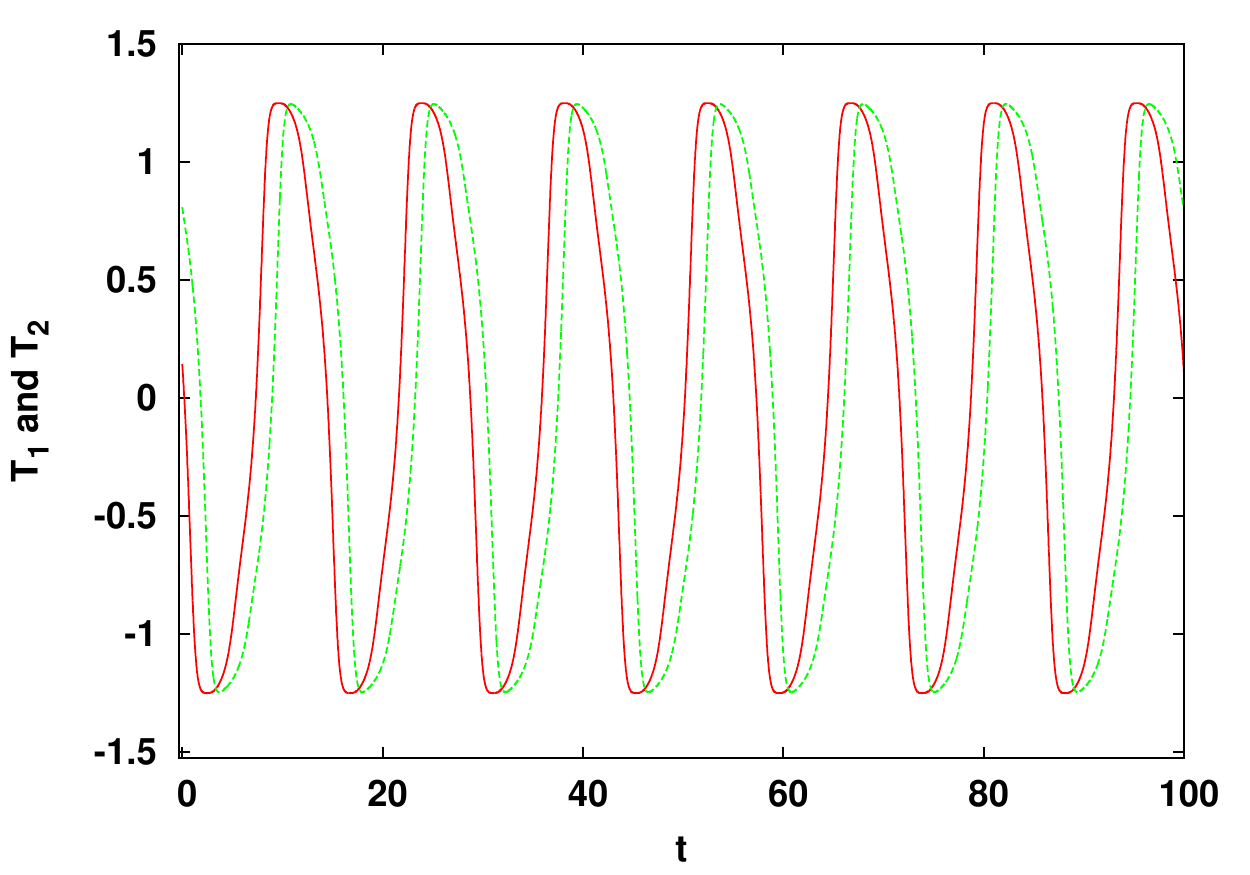} 
			\includegraphics[scale=\SCALE]{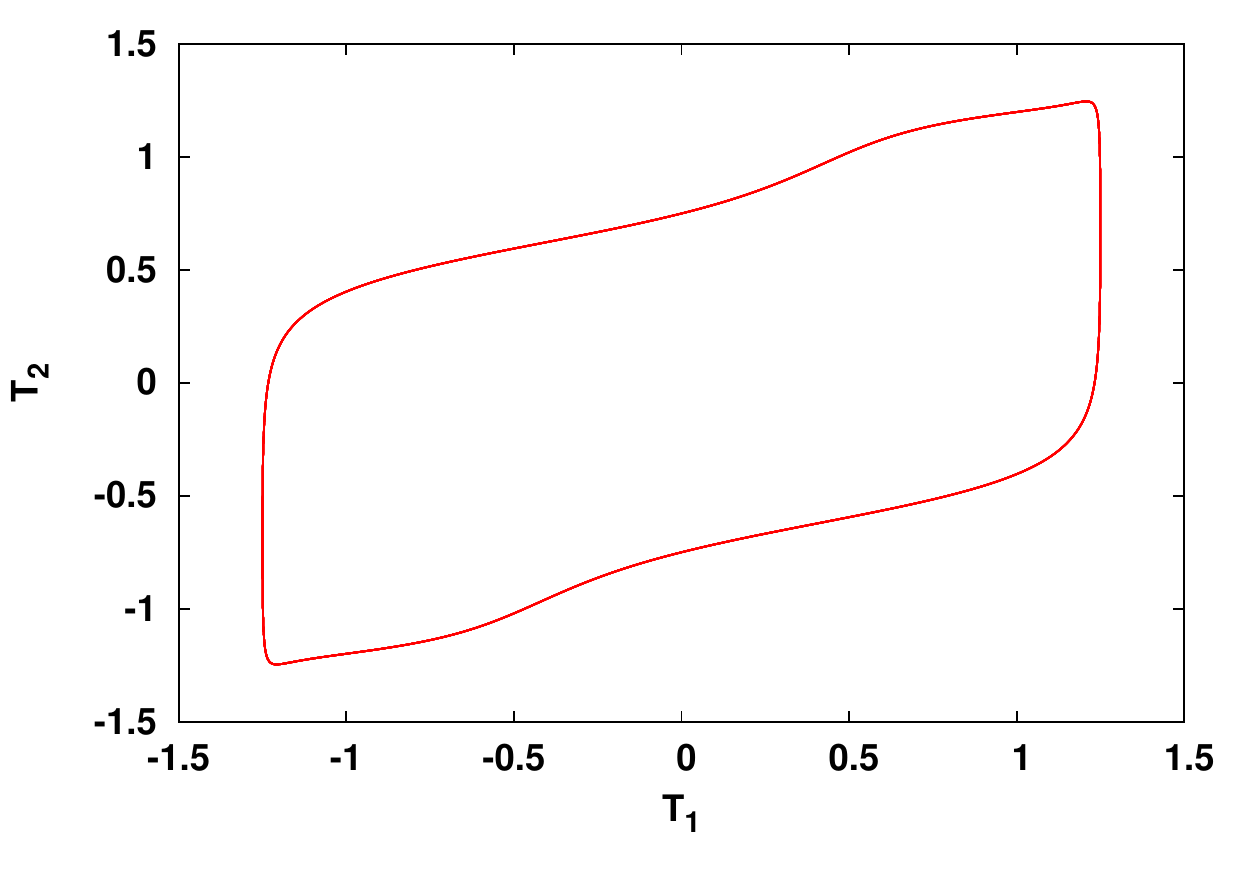}
			(b)
	
			\caption{Temporal evolution of the temperature anomalies of the two sub-regions $T_1$ (in red) and $T_2$ (in green), in the left panels, and the corresponding phase portraits in the $T_1-T_2$ plane in the right panels, for a system with $\alpha_{1} = 0.75, \alpha_{2} = 0.5$, coupling delay $\delta =4$ and inter-region coupling strength  $\gamma$ equal to (a) $0.1$ and (b)  $0.2$, in Eqn.\ref{main}.}	\label{delta_alpha_1}
		\end{figure}

\begin{figure}[H]
			\centering 
			\includegraphics[scale=\SCALE]{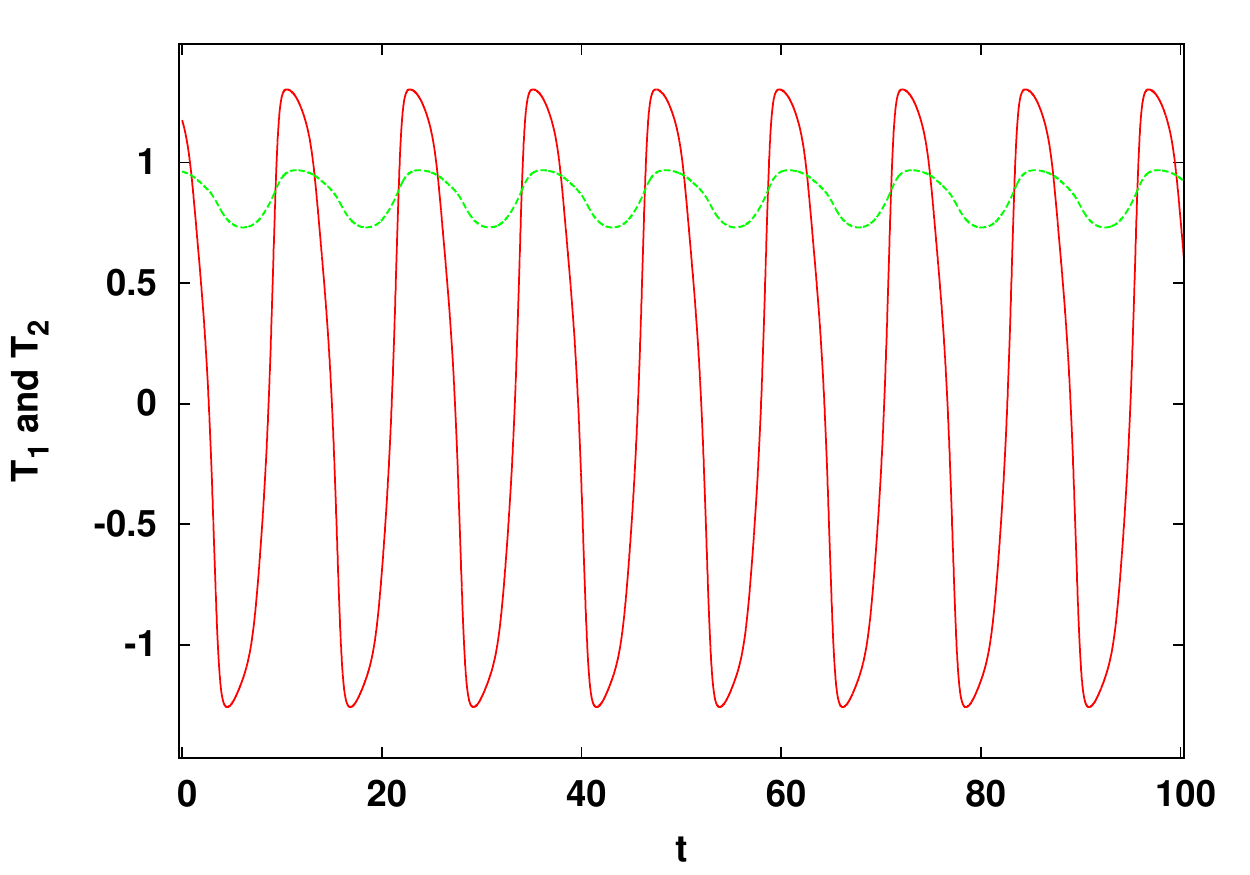}
			\includegraphics[scale=\SCALE]{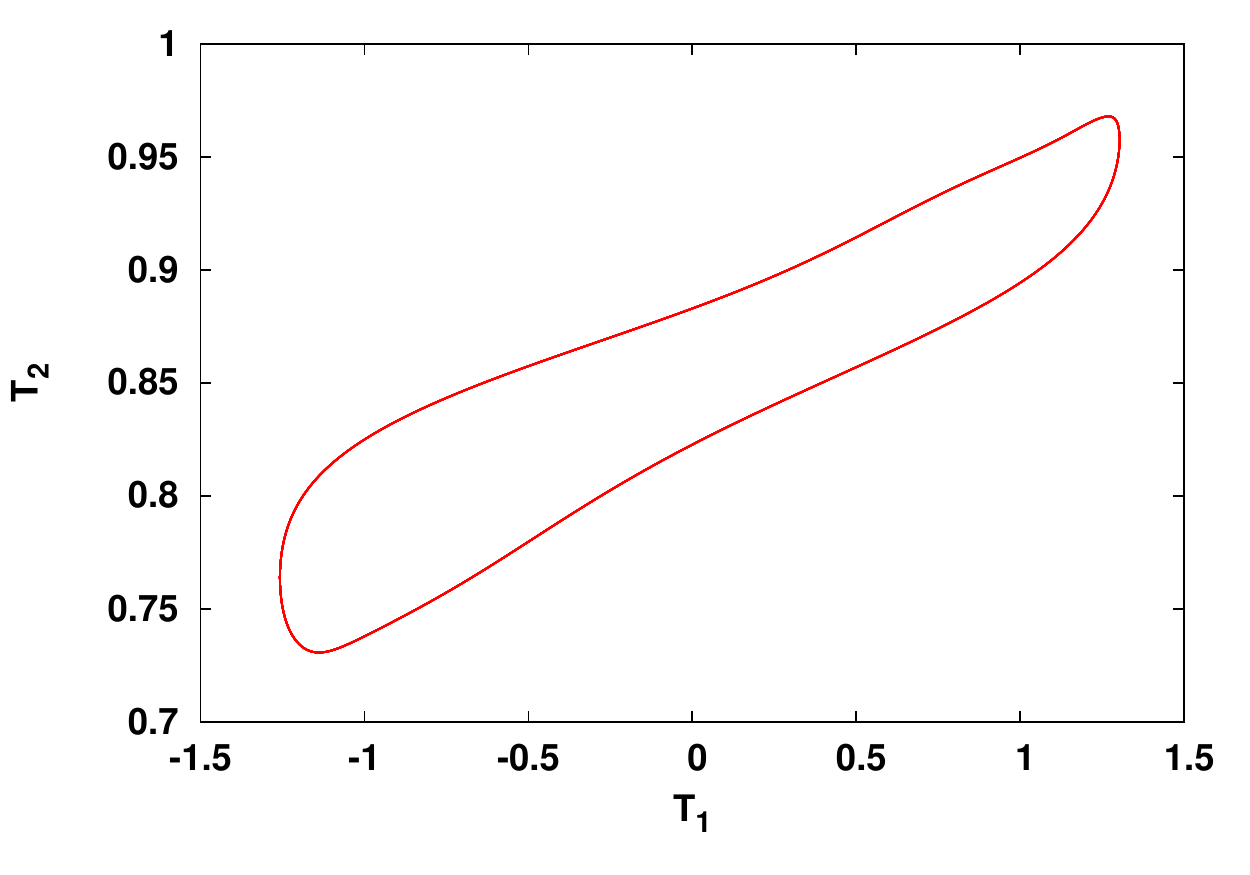}
			(a)
				
			\includegraphics[scale=\SCALE]{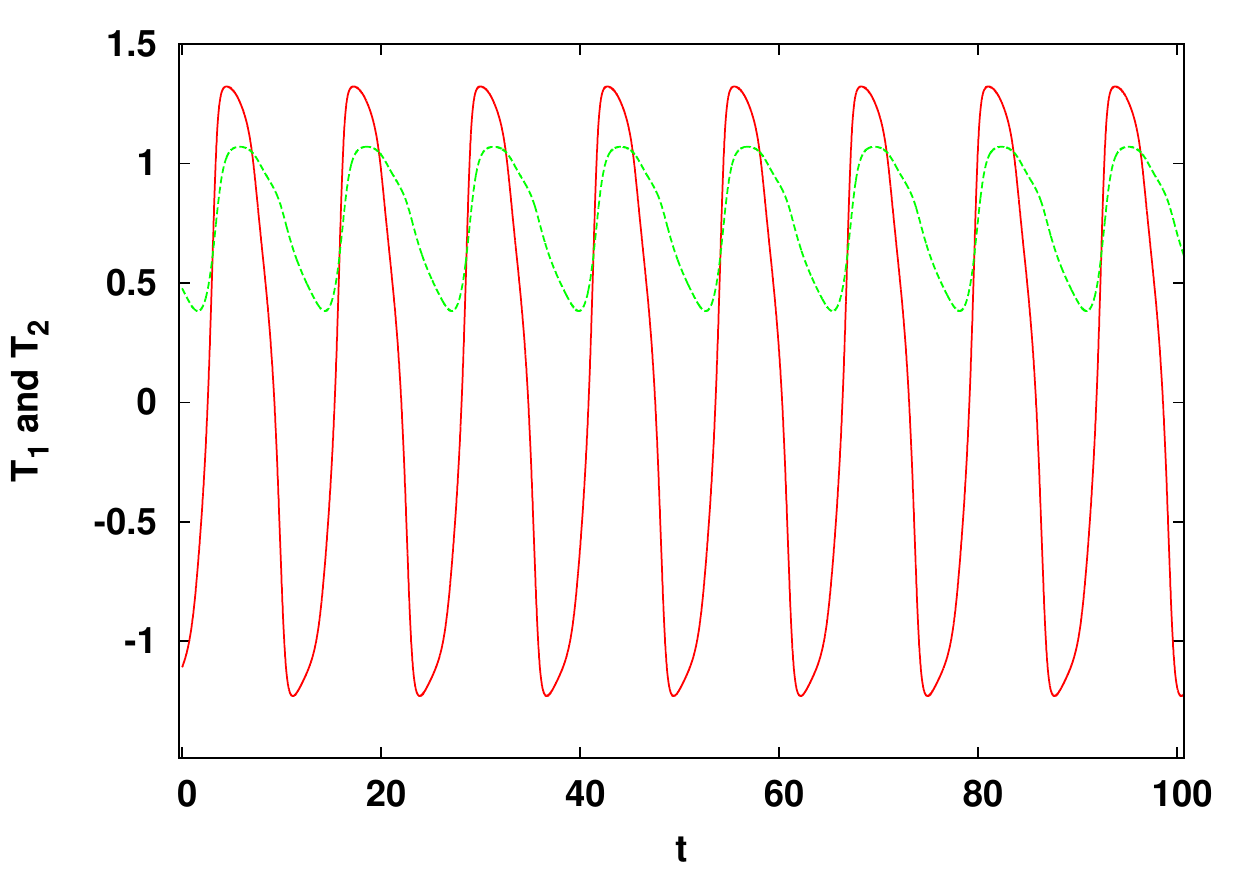} 
			\includegraphics[scale=\SCALE]{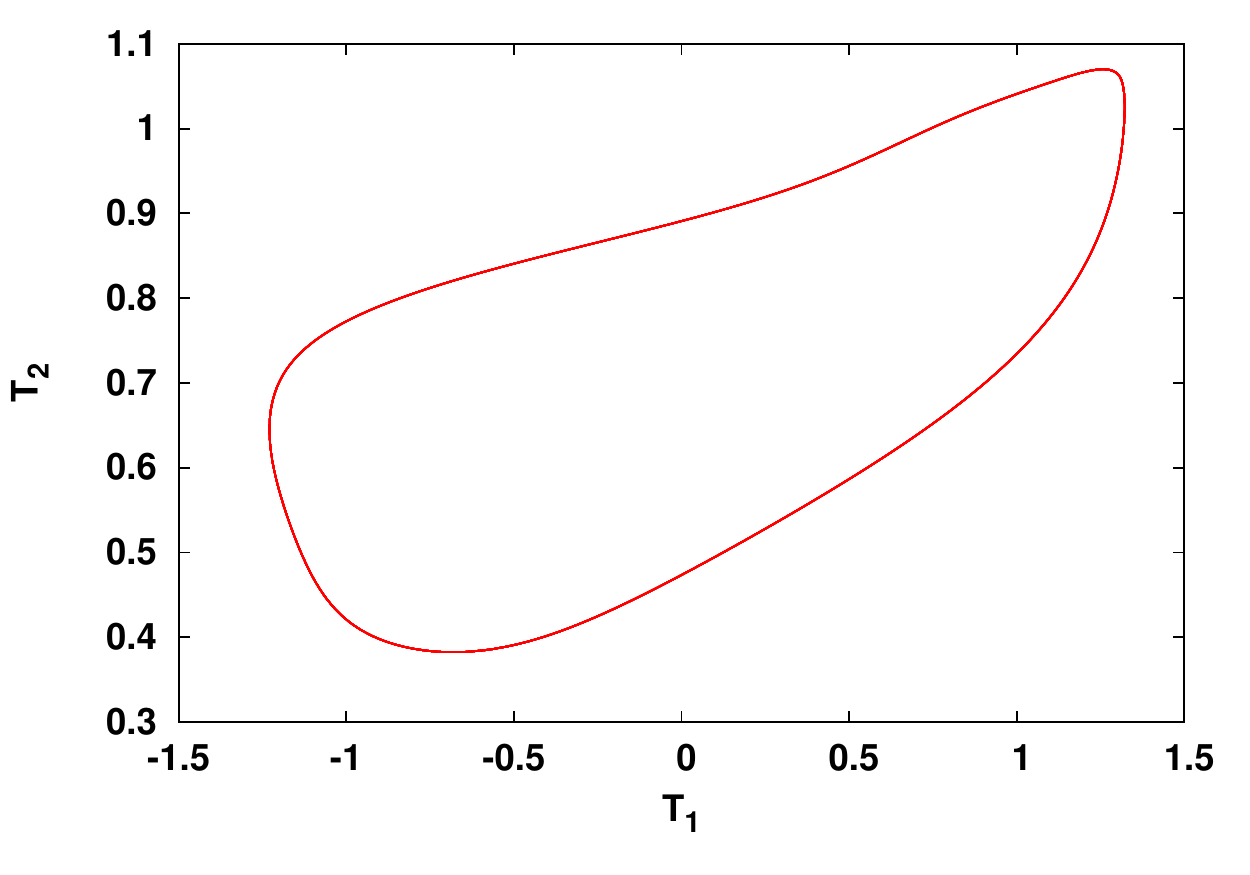} 
			(b)
			\caption{Temporal evolution of the temperature anomalies of the two sub-regions $T_1$ (in red) and $T_2$ (in green), in the left panels, and the corresponding phase portraits in the $T_1-T_2$ plane in the right panels, for a system with $\alpha_{1} = 0.75, \alpha_{2} = 0.25$, delay $\delta =4$ and inter-region coupling strength $\gamma$ equal to (a) $0.1$ and (b) $0.2$, in Eqn.\ref{main}.}\label{delta_alpha_2}		
		\end{figure}

			Now we consider the effects of different strength for the self-delay coupling
			term in the two sub-regions (i.e.  $\alpha_1 \ne \alpha_2$) with the uniform
			delays $\delta_1=\delta_2=\delta$ and inter-region coupling strengths $\gamma$.
			Figs. \ref{delta_alpha_1}-\ref{delta_alpha_2} show the typical dynamics emerging
			under varying differences in the two sub-regions $\Delta \alpha = \alpha_1 - \alpha_2$.
	
		When the difference in the strengths of the self-delay coupling  is small
		($\Delta \alpha < \alpha_{1,2}$), we observe that both sub-regions display similar
	 	behaviour for strong inter-region coupling (cf. Fig. \ref{delta_alpha_1}b ). However 
	    for weaker inter-region coupling, different dynamical behaviour emerges in the two
       sub-regions (cf. Fig. \ref{delta_alpha_1}a ) . Typically, the region with stronger self-delay coupling shows regular
       behaviour, while the region with weaker self-delay coupling shows complex behaviour.
       
       This type of complex oscillation is qualitatively very similar to ENSO observational
       data \cite{Soon2,basicmodel}.
			
	   When the difference in $\alpha$ is large ($\Delta \alpha > \alpha_{1,2}$), then
	   the nature of oscillations in the two sub-regions can be very different. For
	   instance in Fig. \ref{delta_alpha_2} one observes that one sub-region displays 
	   large amplitude oscillations in the temperature anomaly, while the other 
	   sub-region displays very small amplitude oscillations. So we see that 
	   non-uniformity in the self-coupling strengths in the systems can significantly
	   affect the temperature anomaly of mean sea surface temperature in neighbouring 
	    sub-regions.

\begin{figure}[H]
			\centering
			\includegraphics[scale=\SCALE]{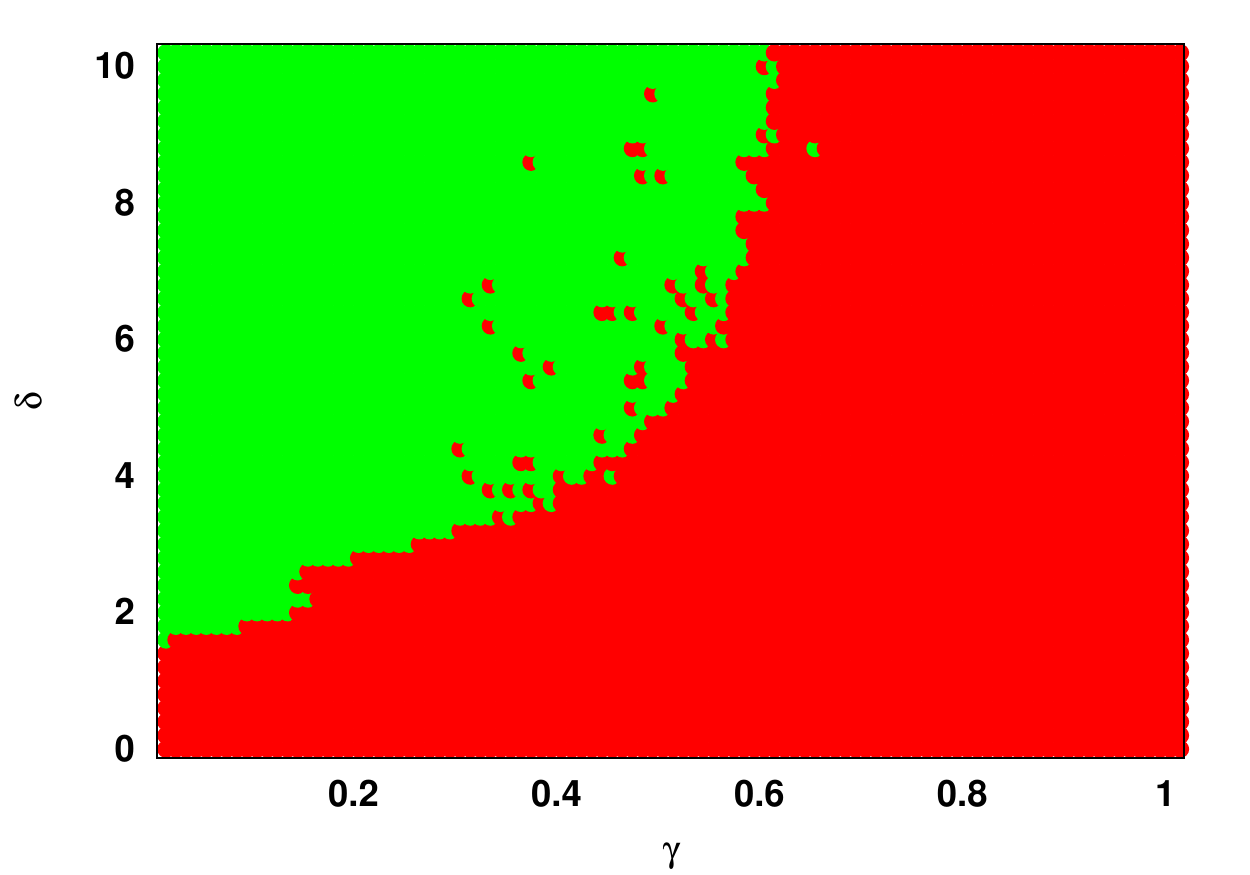}
			\includegraphics[scale=\SCALE]{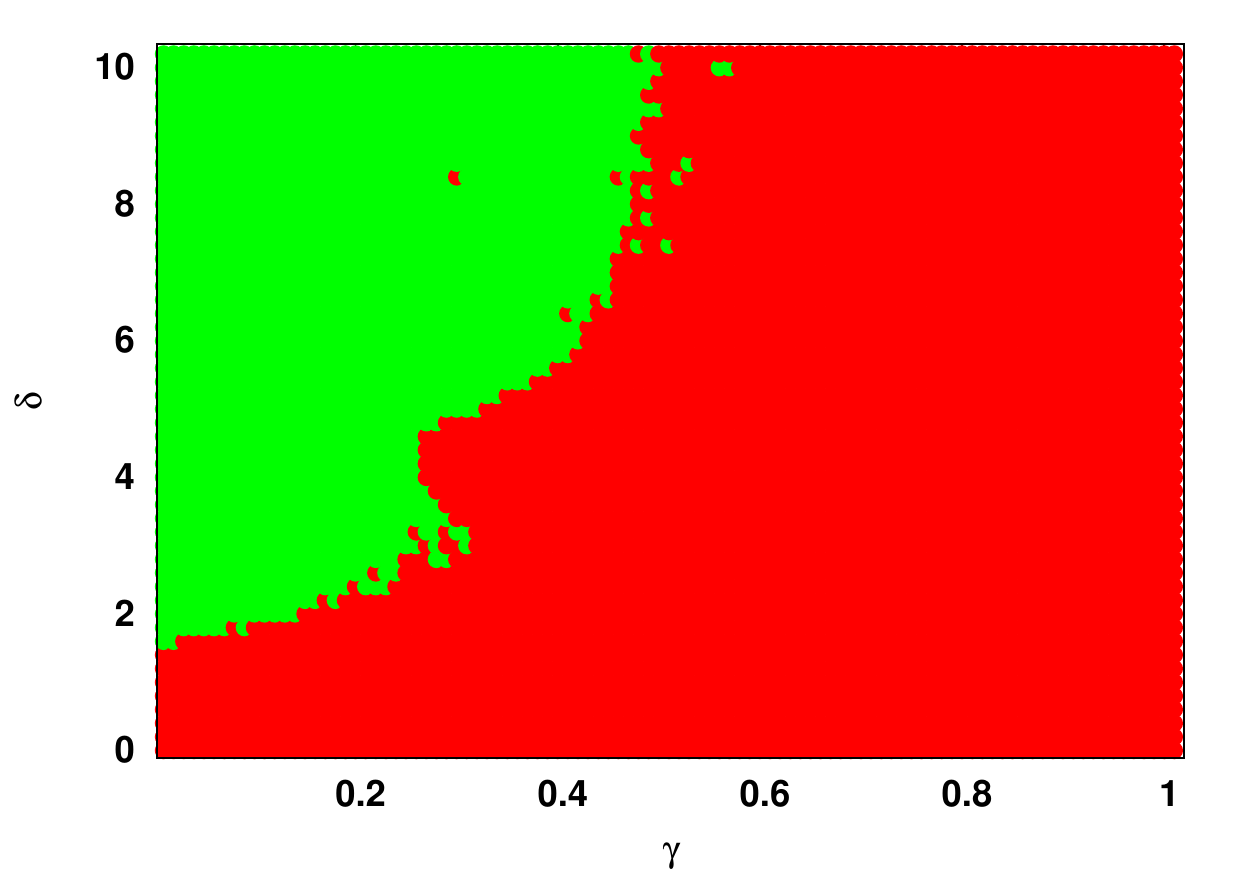}
			\caption{Phase diagram showing the dynamics of the temperature anomaly in mean sea surface temperature ($T_{1}/T_{2}$), with respect to inter-region coupling $\gamma$ ($\gamma > 0 $) and delay $\delta_1=\delta_2=\delta$. Here the strength of delayed coupling in Eqn. \ref{main} is different in the two regions, with $\alpha_{1} = 0.75,  \alpha_{2} = 0.5$ (left) and $\alpha_{1} = 0.75,  \alpha_{2} = 0.25$ (right). The red color represents oscillator death and green represents heterogeneous oscillations. Clearly, oscillator death and heterogeneous oscillations are predominant.}
		   \label{phase_diagram_delta_alpha_1a}
		\end{figure}

\begin{figure}[H]
			\centering 
			\includegraphics[scale=\SCALE]{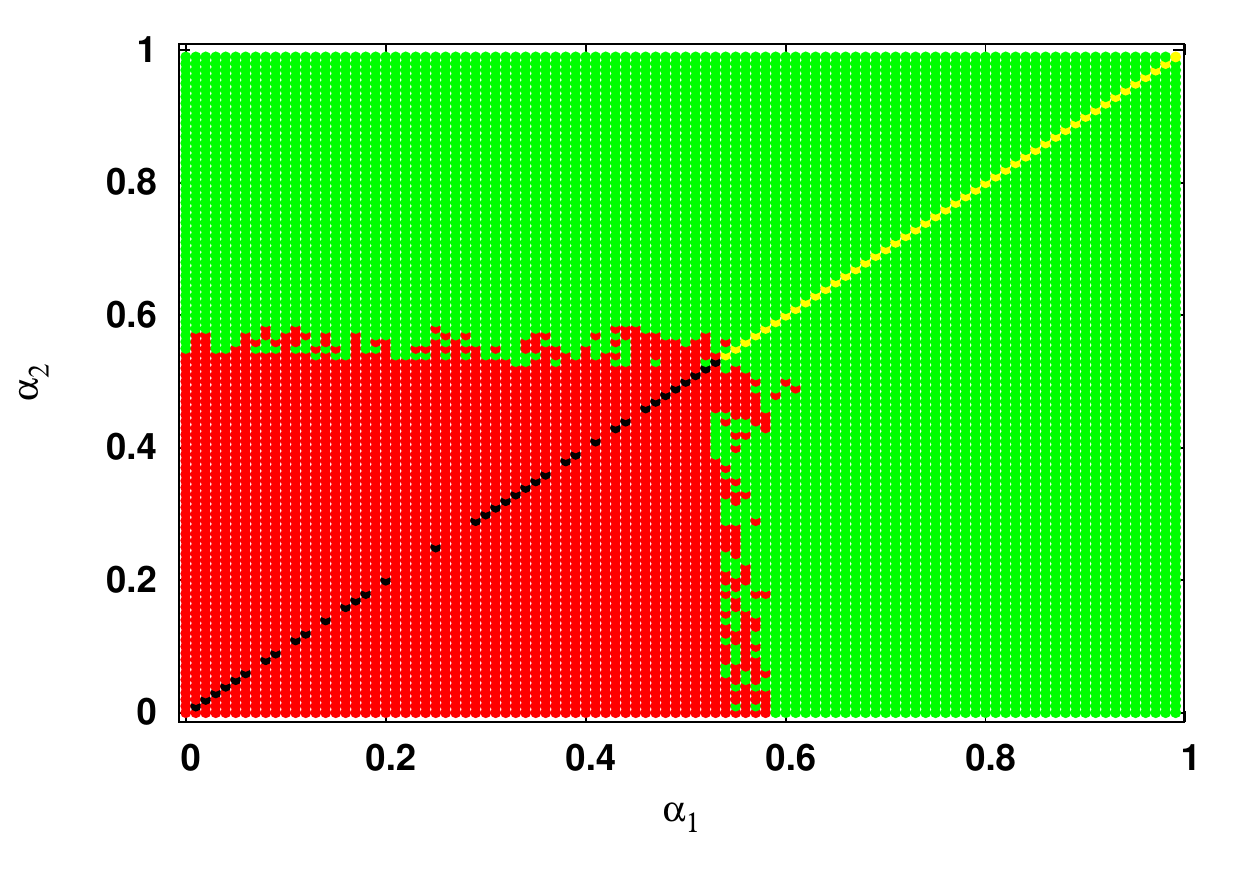}
			\includegraphics[scale=\SCALE]{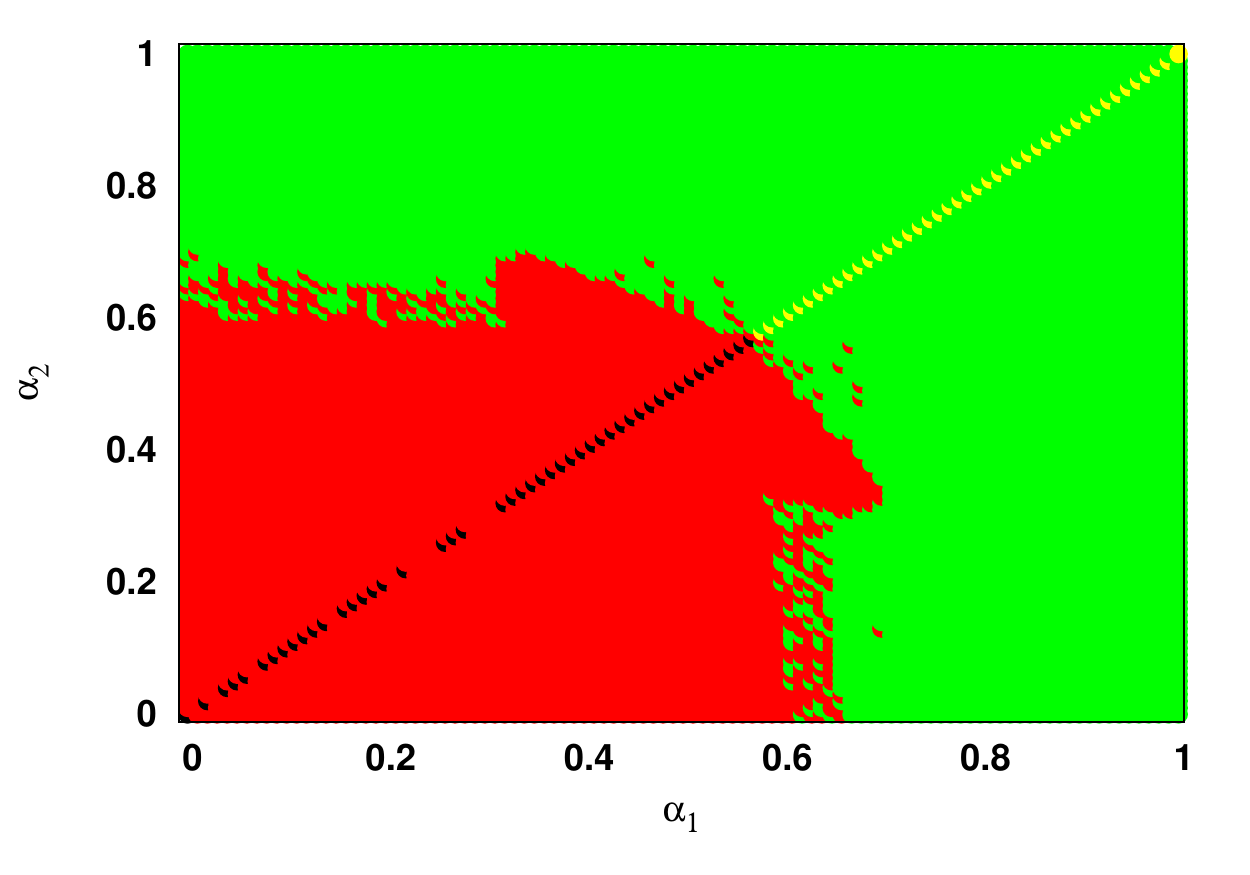}
			\caption{Phase diagram showing the dynamics of the temperature anomaly in mean sea surface temperature of a sub-region ($T_1/T_2$) with respect to self-delay coupling strength $\alpha_{1}$ of first region and self-delay coupling strength $\alpha_{2}$ of second region. The inter region coupling strength $\gamma =0.1$ (left) and $\gamma =0.2$ (right) and delay in the two regions is $\delta_{1}=\delta_{2}=\delta=4$ in Eqn. \ref{main}. The black color represents amplitude death, red represents oscillator death, yellow represents homogeneous oscillations and green represents heterogeneous oscillations. Clearly, oscillator death and heterogeneous oscillations are predominant.}\label{non_identical}
		\end{figure}

			The dependence of the emergence of oscillations on heterogeneity is displayed in more detail in a series of phase diagrams in figures \ref{phase_diagram_delta_alpha_1a} and \ref{non_identical} showing the parameter regimes that yield fixed points and those that gives rise to oscillations in the two sub-regions. Clearly, {\em the parameter region supporting oscillations is larger for weaker inter-region coupling strengths and small difference in self-delay coupling strengths of the two sub-regions}.
			
\begin{figure}[H]
			\centering 
			\includegraphics[width=\twoFigureSize]{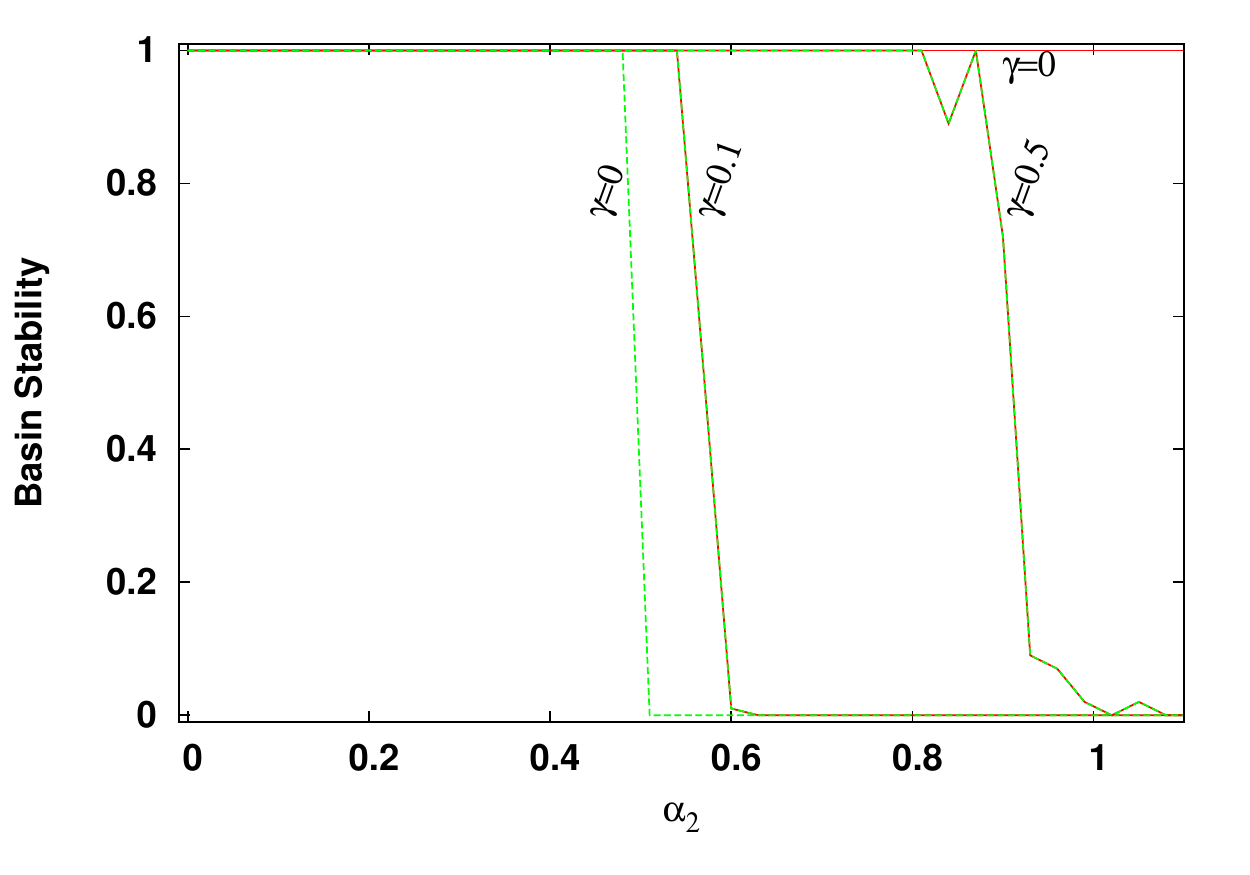}
			\includegraphics[width=\twoFigureSize]{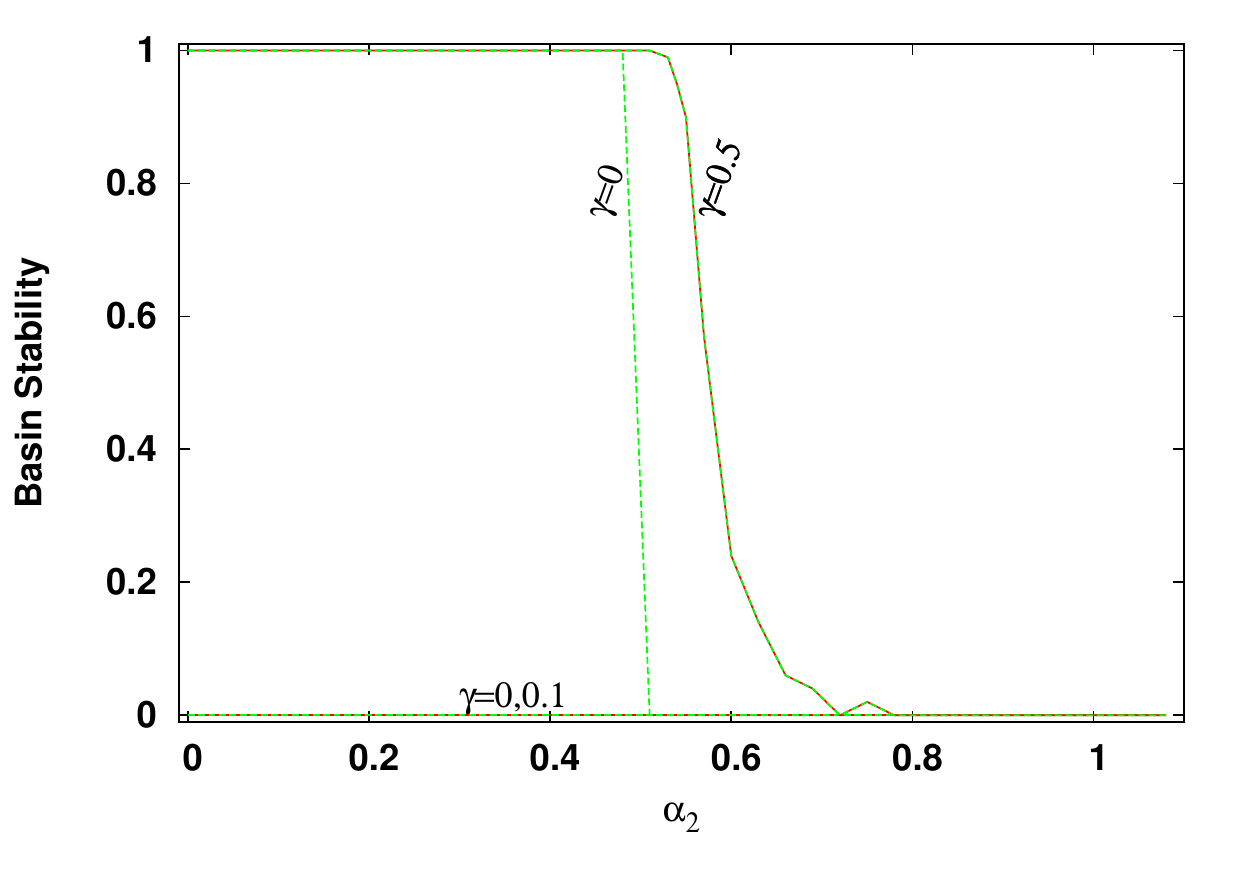}
			\caption{Basin stability of the fixed point state (estimated by the fraction of initial states that evolve to a steady state), as a function of $\alpha_{2}$. The red color represents region 1 and green color represents region 2. Here delay $\delta =4$, and inter-coupling strength $\gamma = 0$, $0.1$ and $0.5$  in  Eqn. \ref{main}, for (left) $\alpha_{1}=0$ and (right) $\alpha_{1}=0.75$.}
			\label{basin_fp_delta_alpha}
		\end{figure}

			We estimate the basin stability for the fixed point state, by finding the fraction of initial conditions that evolve to fixed points. If this fraction is one, the fixed point state is the global attractor of the dynamics. When this fraction is zero, none of the sampled initial conditions evolve to fixed points, and the system goes to an oscillatory state instead. When the fraction is larger than zero and less than one, we have co-existence of attractors (namely certain initial conditions evolve to fixed points, while others yield oscillations). The basin of attraction, as a function of $\alpha_{2}$, keeping $\alpha_1$ fixed, is displayed in Fig. \ref{basin_fp_delta_alpha}, for different values of inter-region coupling strengths.  It is clearly seen that the region of co-existence of fixed points and oscillations is narrower for lower inter-region coupling, and wider for higher inter-region coupling strengths. Thus it is a evident that strong inter-region coupling $\gamma$ favours larger parameter regions of oscillation suppression, and also yields a larger parameter range where fixed points states co-exist with oscillatory states.

\begin{figure}[H]
			\centering 
			\includegraphics[scale=\SCALE]{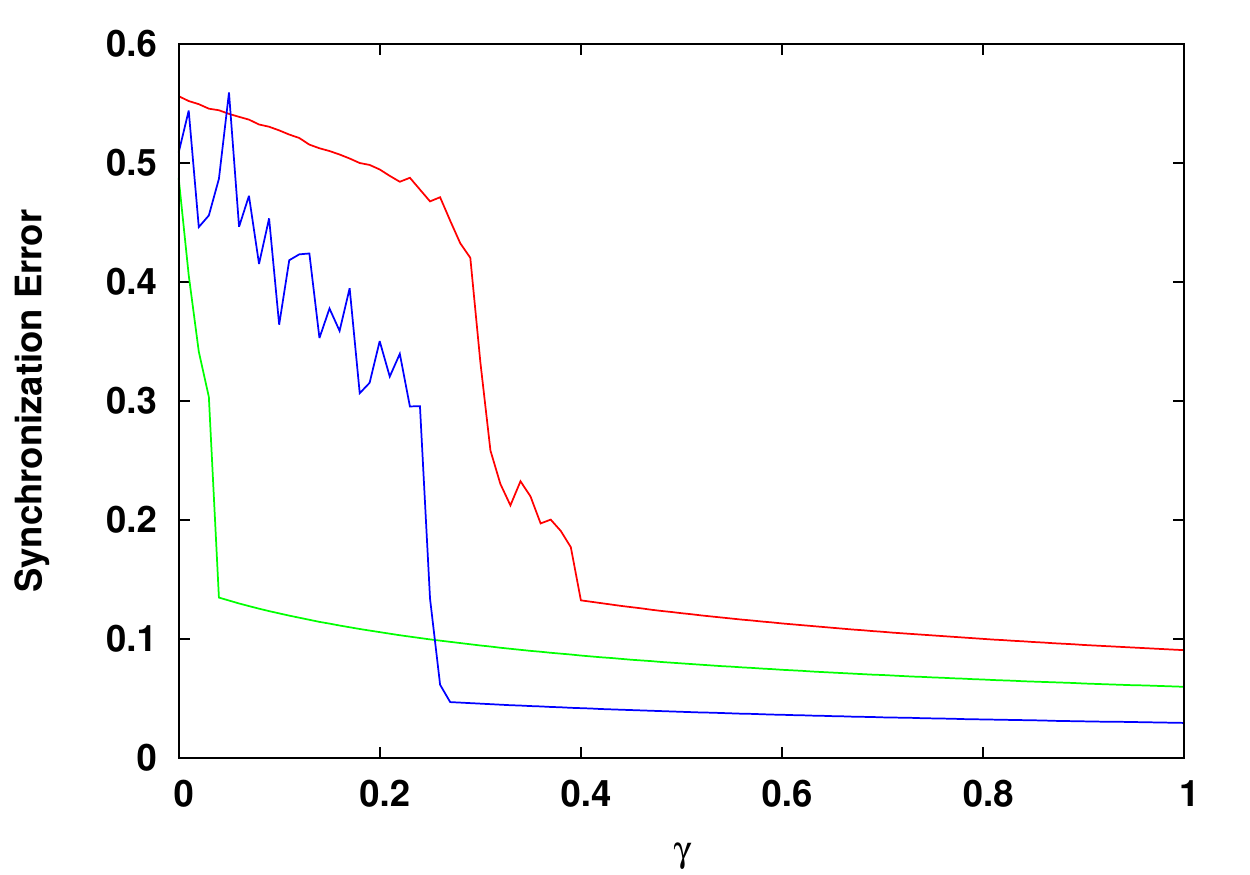}
			\caption{Synchronization error in a coupled system with self-delay coupling strength $\alpha_{1}=0$ and $\alpha_{2}= \alpha_{1} + \Delta\alpha$ in Eqn. \ref{main}, as a function of inter-region coupling strength $\gamma$, for $\alpha_{2} = 0.75$ (red) $\alpha_{2}=0.5$ (green) and $\alpha_{2}=0.25$ (blue).}
			\label{sync1}
		\end{figure}

		We next investigate the synchronization properties of the two sub-regions. In order to quantitatively assess the degree of synchronization we calculate the synchronization error, namely the difference $|T_1-T_2|$ averaged over time and over different initial conditions. Fig. \ref{sync1} shows this quantity by varying inter-region coupling strengths $\gamma$.
		
		\begin{figure}[H]
			\centering 
			\includegraphics[scale=\SCALE]{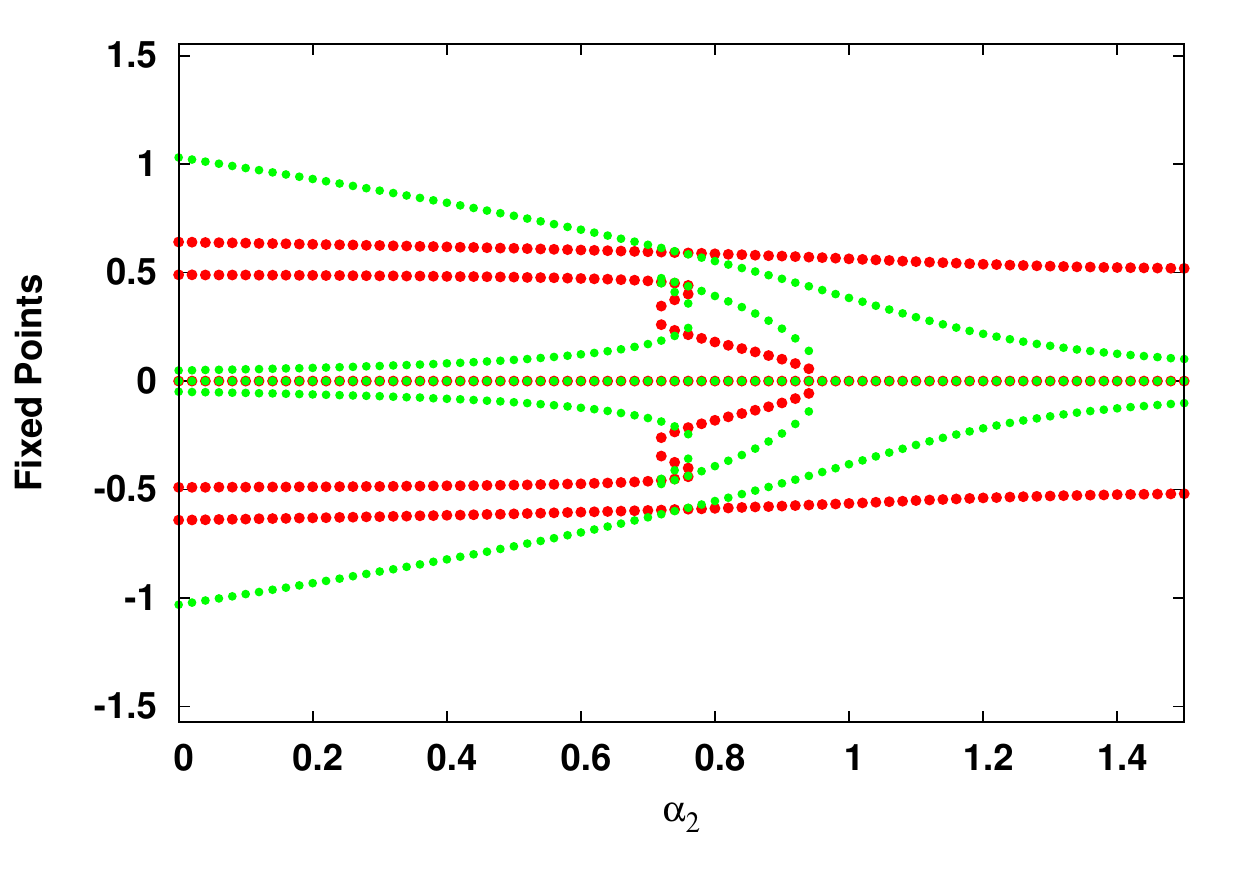}
			\includegraphics[scale=\SCALE]{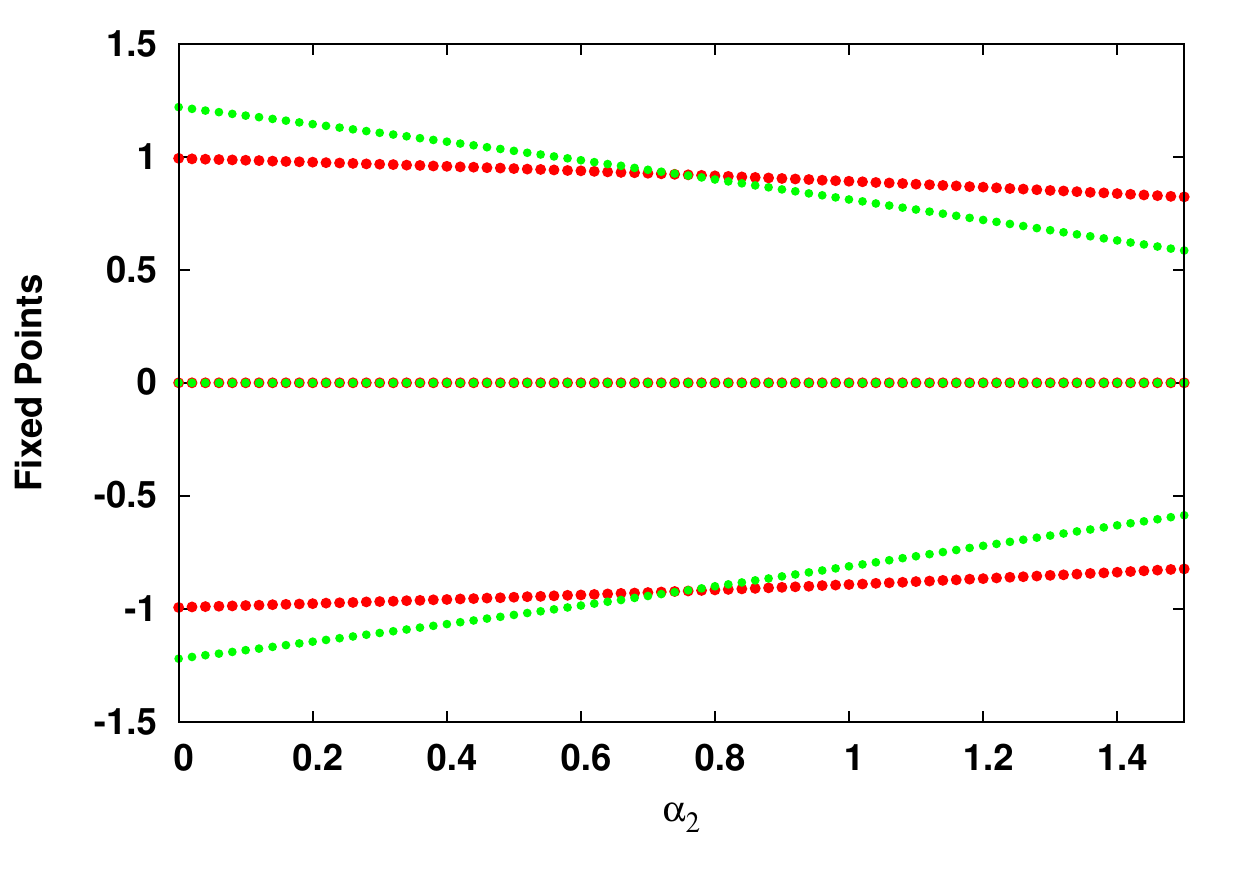}
			\caption{Fixed point solutions of $T_1$ and $T_2$ arising from Eqn. \ref{small_delay} (red for region $1$ and green for region $2$) versus strength of the self-delay coupling $\alpha_2$ of region 2, with $\alpha_1 = 0.75$, and inter-region coupling $\gamma$ equal to $0.1$ (left) and $0.6$ (right).}
			\label{stability_analysis_hetero}	
				
		\end{figure}
		{\em Analysis: } We find a richer pattern of fixed point solutions for the heterogeneous case (cf. Fig. \ref{stability_analysis_hetero}), as compared to the homogeneous case (cf. Fig. \ref{stability_analysis}), as the solutions $T_1$ and $T_2$ may now be different. For weak inter-region coupling, such as $\gamma=0.1$, $T_1$ has $5$ fixed points for $0 \le \alpha_2 \le 0.7$, of which $2$ are stable nodes, $2$ are saddle points and $1$ is an unstable node. For $0.7 < \alpha_2 < 0.78$ we obtain $9$ fixed points, of which $4$ are stable nodes, $4$ are saddle points and $1$ is an unstable nodes. For $T_2$ again we get $5$ fixed points for $0.78 \le \alpha_2 \le 0.94$, of which $2$ are stable nodes, $2$ are saddle points and $1$ is an unstable node, and for $0.94 < \alpha \le 1.5$ we get $3$ fixed points in which $2$ are stable node and $1$ is saddle node. The fixed points have different values in the two regions, except for the unstable node. For strong inter-region coupling, such as $\gamma=0.6$, we get $3$ fixed points for $T_1$ and $T_2$, of which $2$ are stable nodes and $1$ is a saddle. In this case too the stable nodes have different values in the two regions, while the saddle points occur at the same value.
	\section{Effect of different delays in the sub-regions}
\begin{figure}[H]
			\centering 
				\includegraphics[scale=\SCALE]{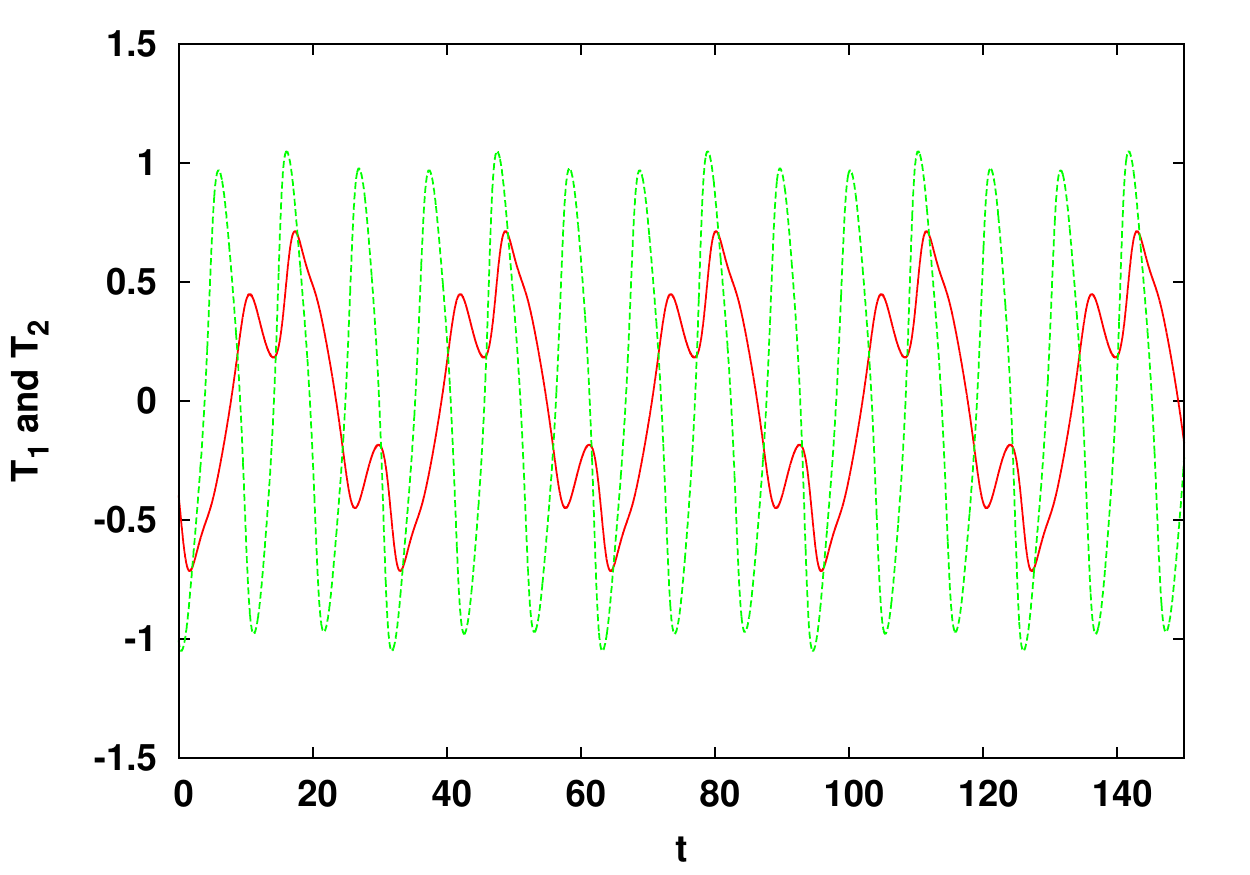}
				\includegraphics[scale=\SCALE]{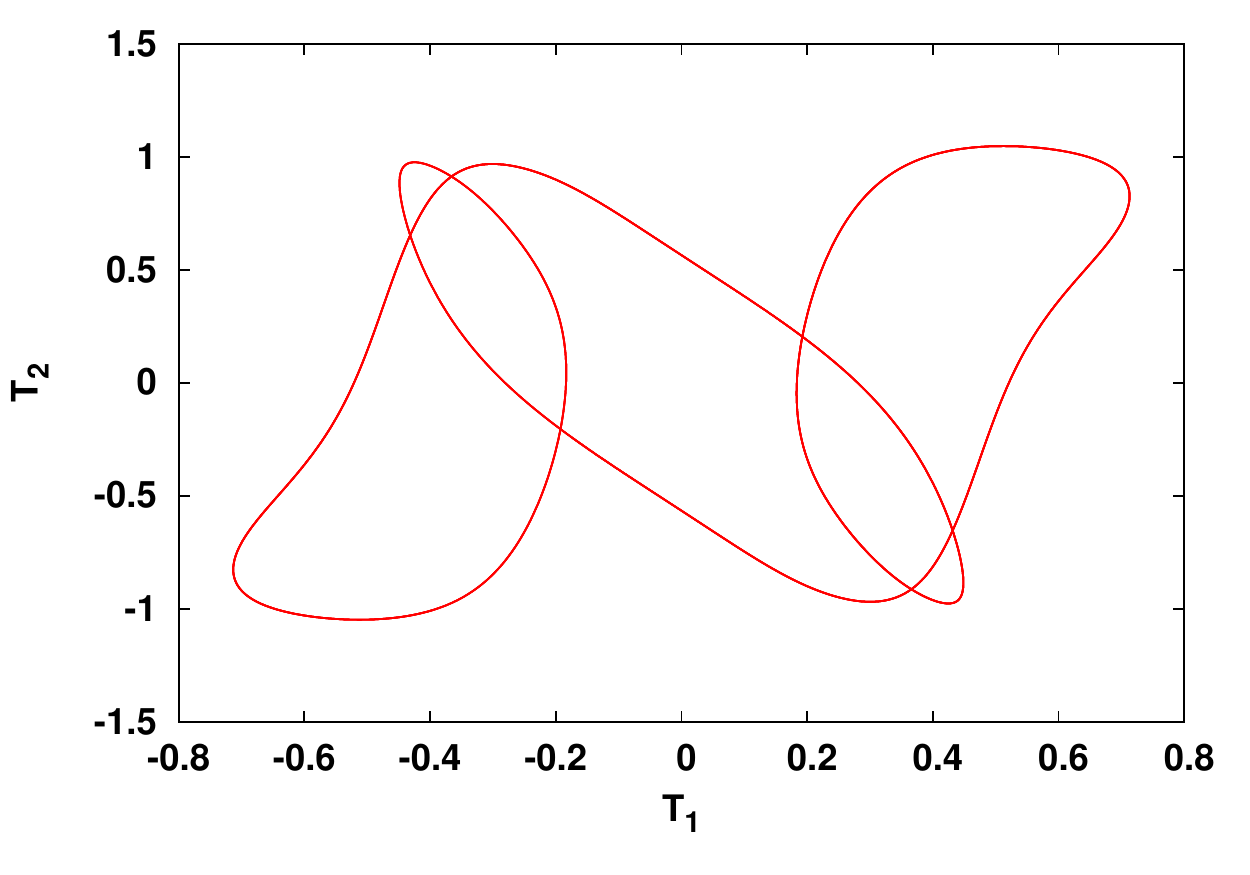} (a)

				\includegraphics[scale=\SCALE]{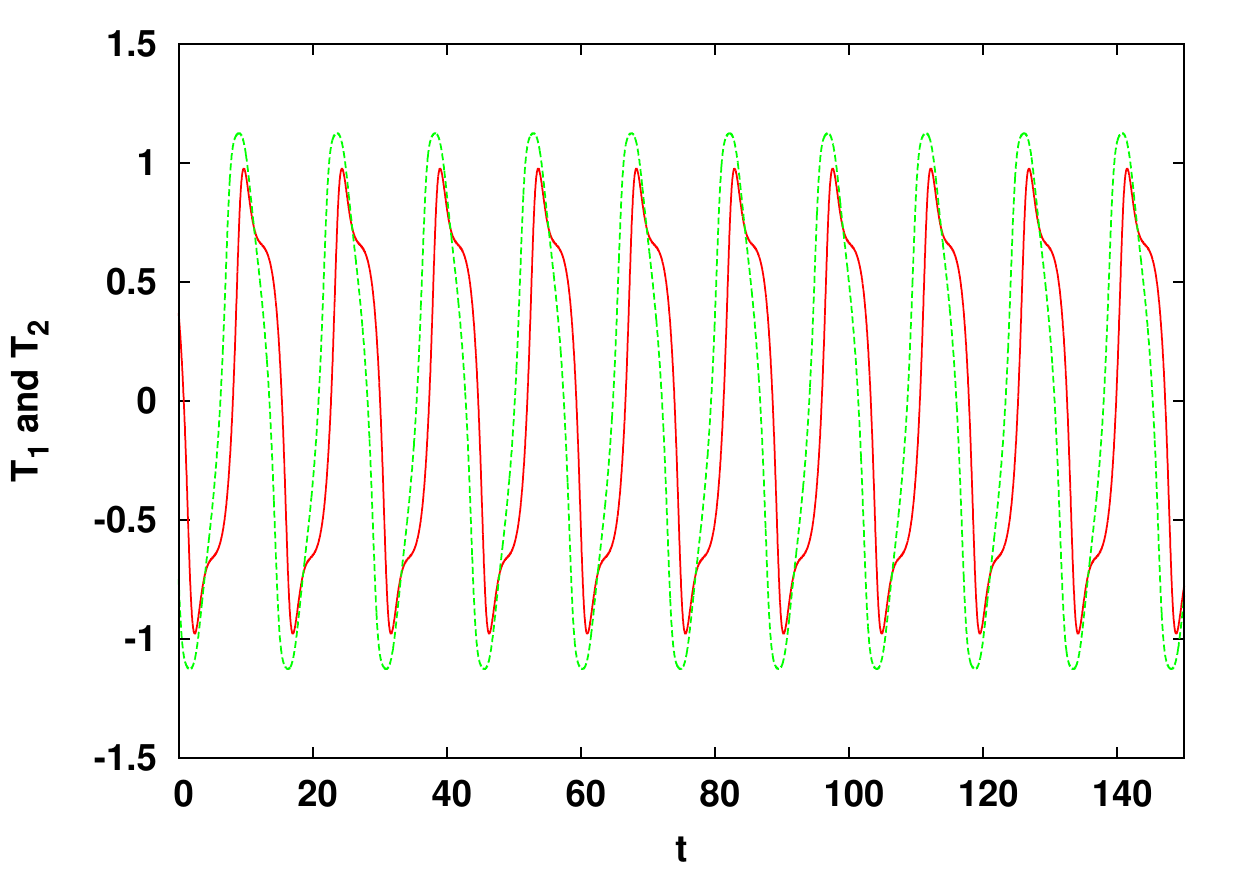}
				\includegraphics[scale=\SCALE]{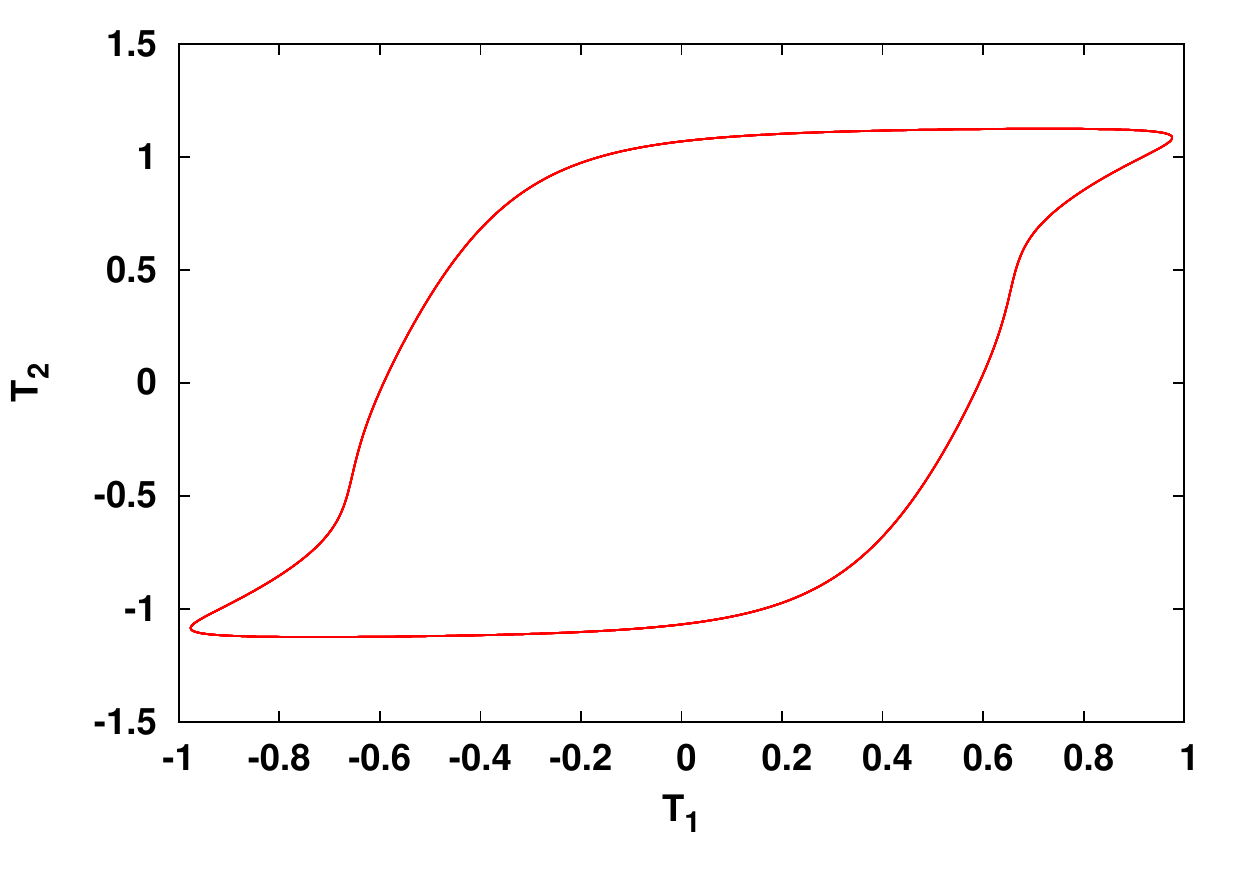} (b)

				\includegraphics[scale=\SCALE]{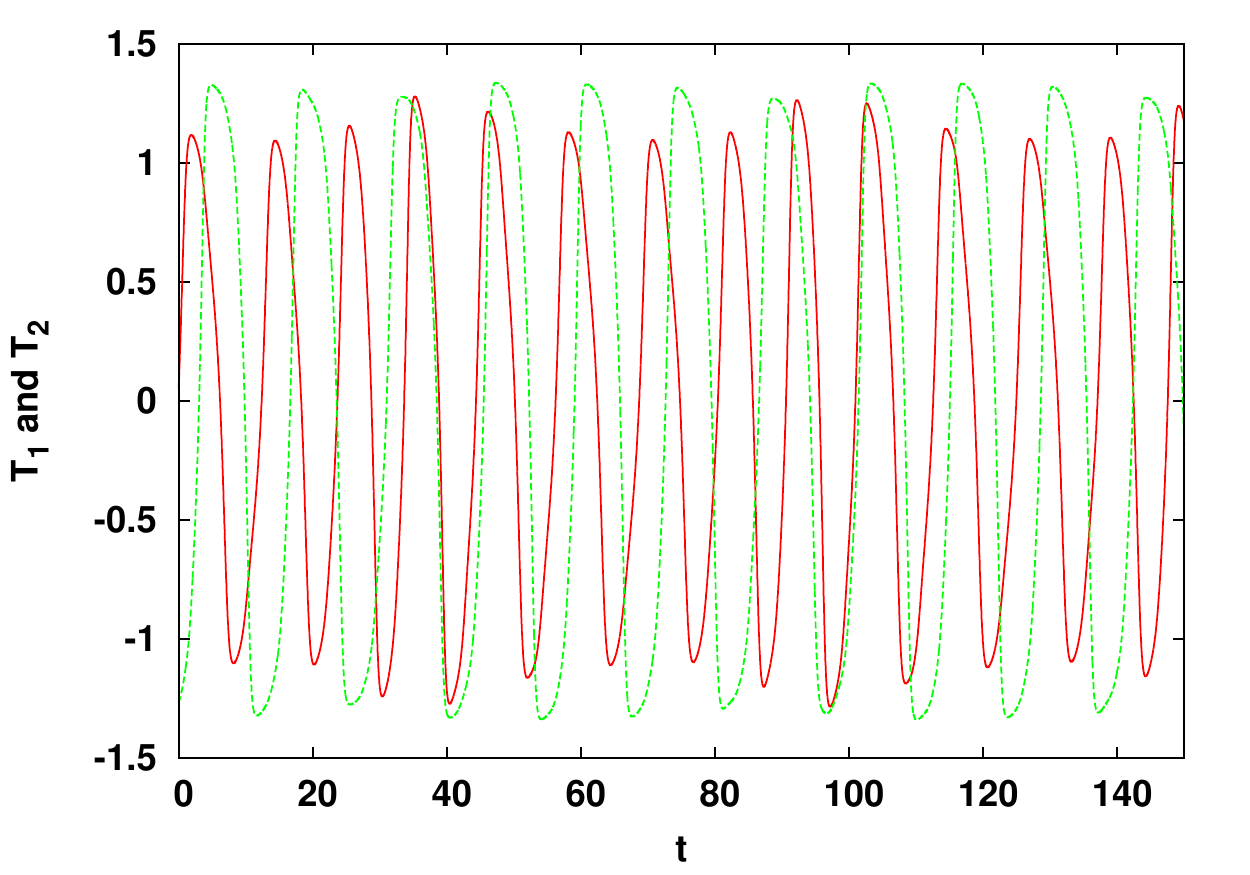}
				\includegraphics[scale=\SCALE]{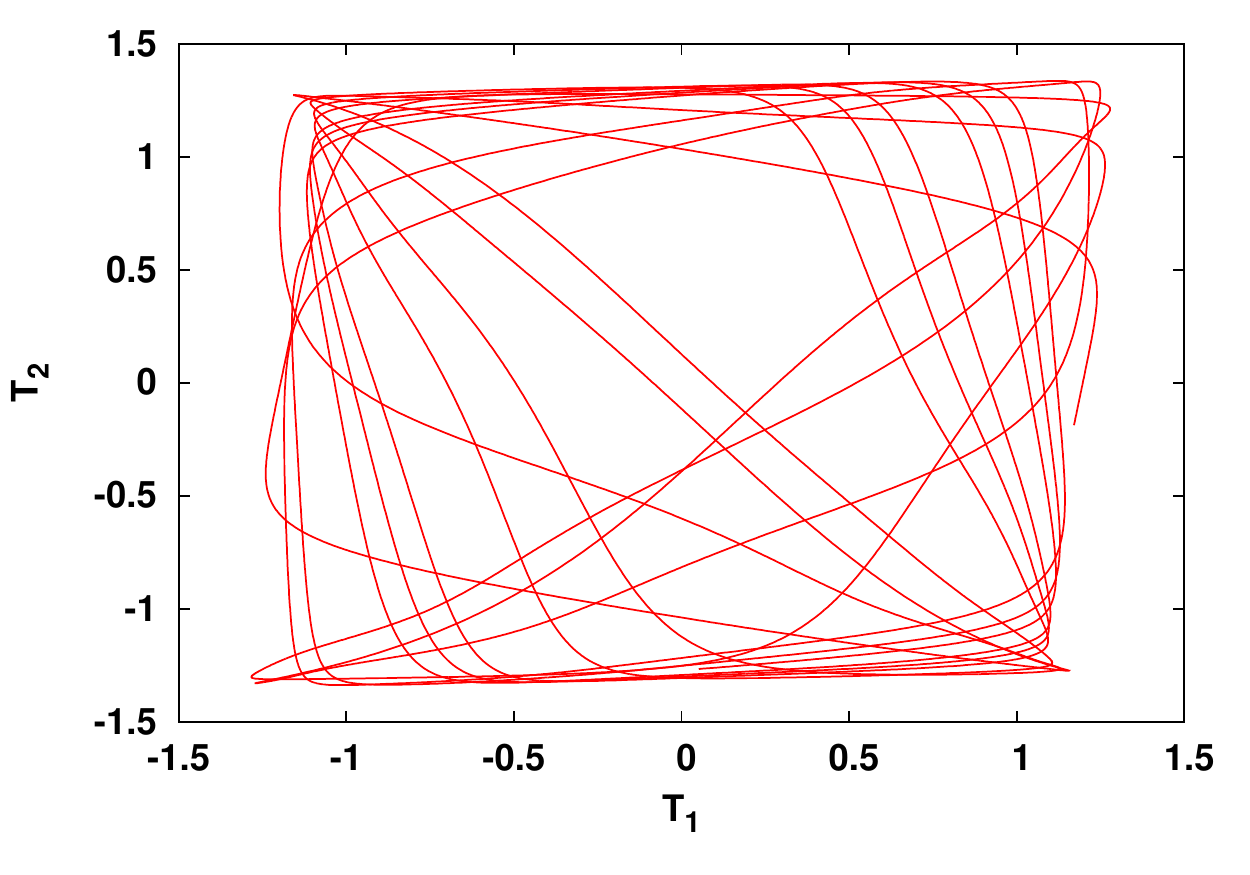} (c)
			\caption{Time evolution of the temperature anomalies of the two sub-regions, with $\alpha_{1}=\alpha_{2}=0.75$ (a) $\delta_{1} = 1$, $\delta_{2} = 2$ and $\gamma=0.1$; (b) $\delta_{1} = 3$, $\delta_{2} = 5$ and $\gamma=0.1$; (c)  $\delta_{1} = 1$, $\delta_{2} = 3$ and $\gamma=0.3$;. The temperature anomaly of region 1, $T_{1}$, is shown in red and for region 2, $T_{2}$ is shown in green. The corresponding phase portrait is displayed on the right panel.}
			\label{patterns}
		\end{figure}

\begin{figure}[H]
			\centering 
				\includegraphics[width=\twoImageSize]{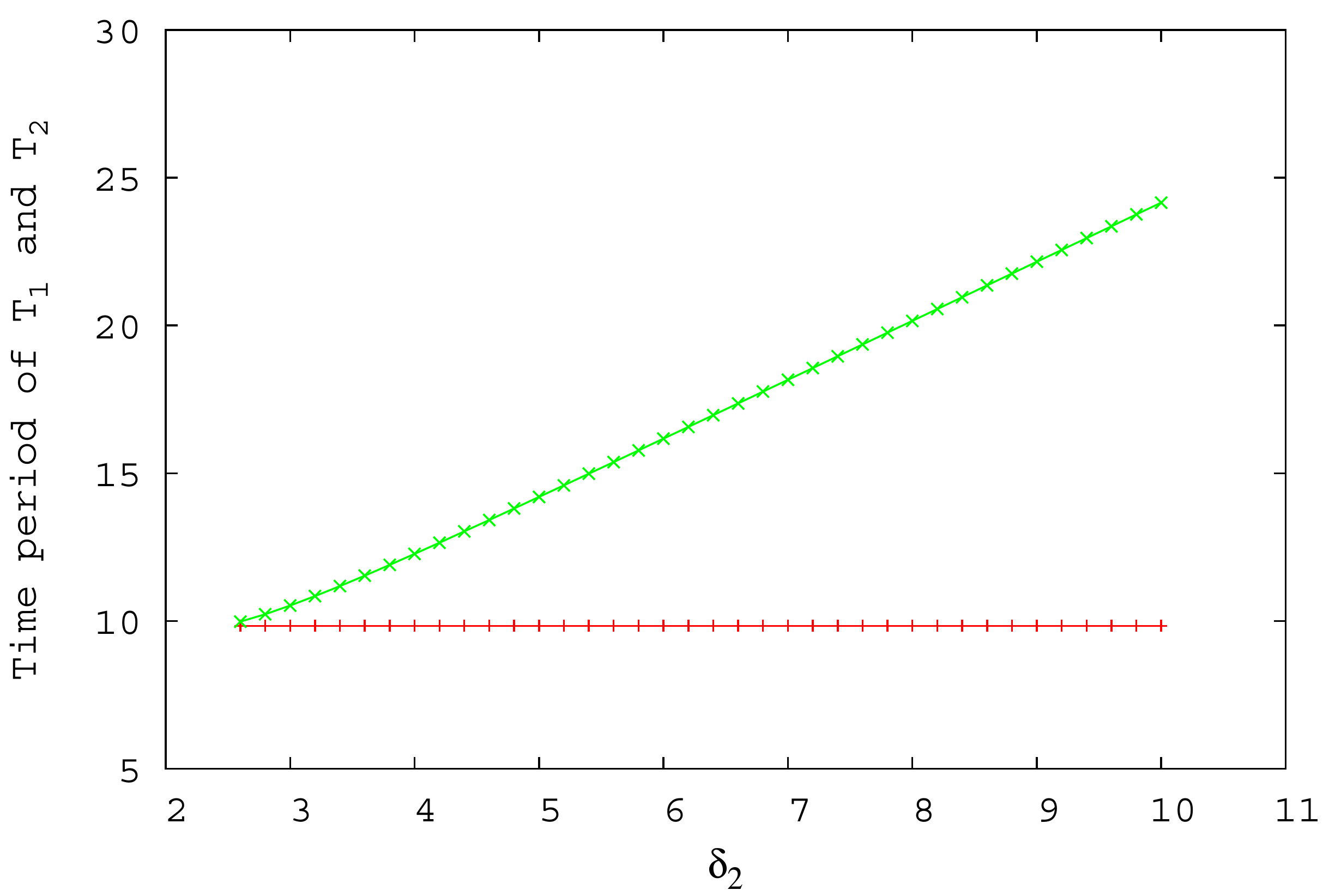}
				\includegraphics[width=\twoImageSize]{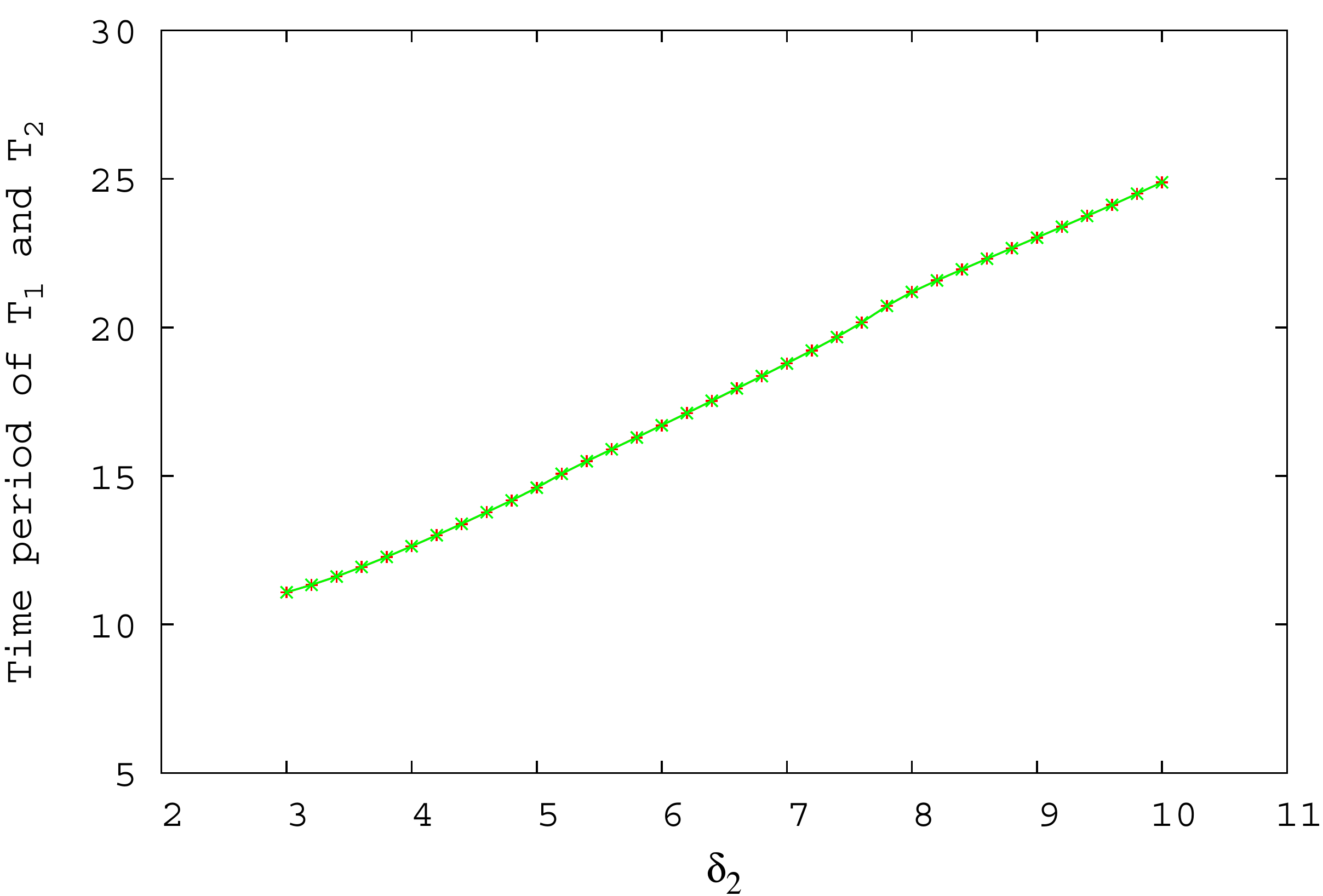}
				\caption{The time period of the temperature oscillations in the two sub-regions $T_{1}$ (red) and $T_{2}$ (green) for $\gamma=0$,  $\delta_{1}=0$ (left) and $\gamma=0.2$ and $\delta_{1}=0$, as a function of the delay $\delta_2$ of the second sub-region. Here $\alpha_{1}=\alpha_{2} = 0.75$.}
				\label{timeperiod}
		\end{figure}

When  the two sub-regions are uniform, the oscillations in temperatures of the sub-regions exhibit simple regular patterns, for all delays. However, when the self-delays are different, with $\delta_{1} \ne \delta_{2}$, complex oscillatory patterns arise. These complex patterns are also qualitatively similar to the actual observations of the ENSO phenomena. Representative examples of these are shown in Fig. \ref{patterns}. Interestingly, though the waveforms of the oscillations in the two sub-regions are different, the time period of the emergent oscillations is the same for both regions. 
The time period of the oscillation increases with increasing delays, as evident from Fig. \ref{timeperiod}. As mentioned above, irrespective of the magnitude of self-delay, the period of oscillation for both regions is the same, as is clearly seen from the overlapping curves for the two sub-regions in Fig. \ref{timeperiod}. 
		Now, in principle, one can compare the oscillation period obtained from Fourier Transforms of SST time series observed in different regions to that obtained from this model, thereby potentially allowing connections between this model and observations. For instance, we can consider two regions along the equator, where the first region extends from $90\degree$ West to $150\degree$ West with the mid-point being $170\degree$ West and the second region extends from $150\degree$ West to $160\degree$ East with the mid-point being $175\degree$ West. The western Pacific boundary is at $120\degree$ East. This gives angular separation of $120\degree$ and $65\degree$ of longitude for the waves to travel, for the two regions respectively, and corresponds to a distance $120(2\pi/360)\times r_{Earth} = 13.35\times10^6m$ and $65(2\pi/360)\times r_{Earth}=7.23\times10^6m$ for the two regions, where $r_{Earth}=6.37\times10^6 m$. Thus delays for the Rossby waves are $329$ days and $178$ days, and delays for the Kelvin waves are $110$ days and $59$ days in the two regions, and the transient time ($\Delta$)  taken by these waves are $439$ days and $237$ days respectively.
		\newpage
		 So the actual time period of the oscillations can be obtained from the time series arising from the dimensionless Eqn.~3, by a scaling factor of $1/k$, where $k= \delta/\Delta$, with $k$ being different in non-identical sub-regions. The value thus obtained can, in principle, be compared to the oscillation period obtained from observations in these different regions. This suggests a manner in which to obtain potential connections between this model and observations.
		
\begin{figure}[H]
				\centering 
				\includegraphics[scale=\SCALE]{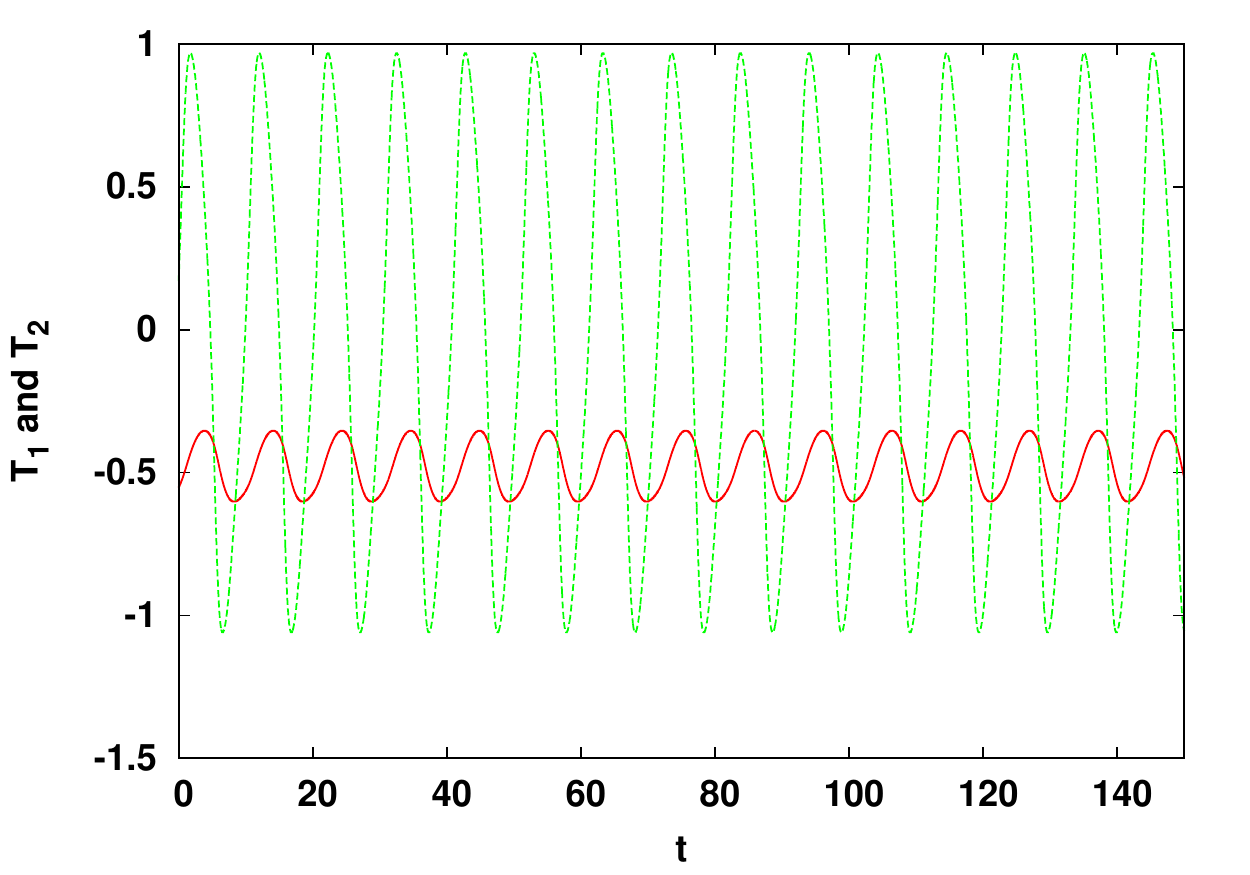}
				\caption{Time evolution of the temperature anomalies of the two sub-regions, with self-delay $\delta_{1} = 0$ in region 1 and $\delta_{2} = 2$ in region 2. The inter-region coupling strength is $\gamma = 0.1$ and self-delay coupling strength is $\alpha=0.75$. The temperature anomaly of region 1, $T_{1}$, is shown in red and for region 2, $T_{2}$ is shown in green.}\label{fp1}
		\end{figure}

\begin{figure}[H]
				\centering 
				\includegraphics[scale=\SCALE]{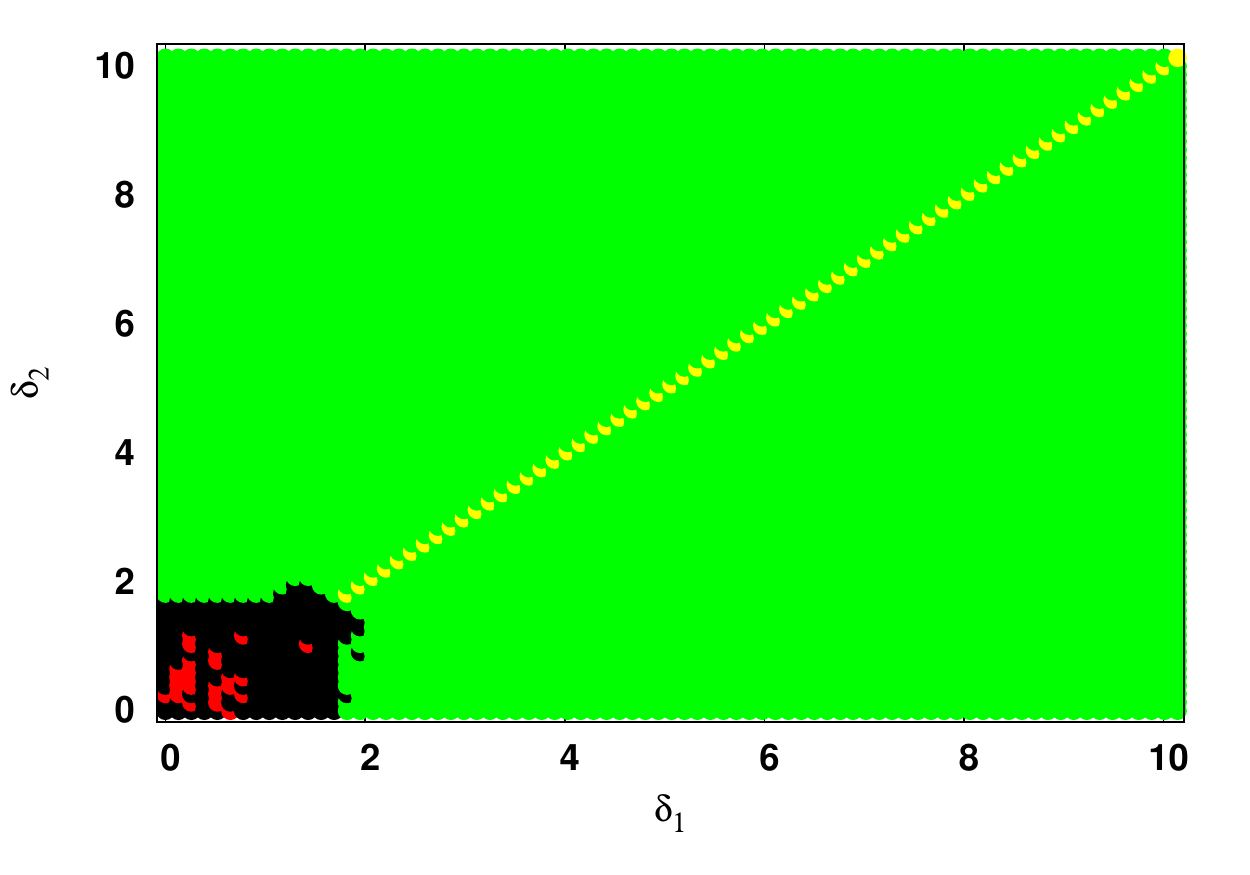}
				\includegraphics[scale=\SCALE]{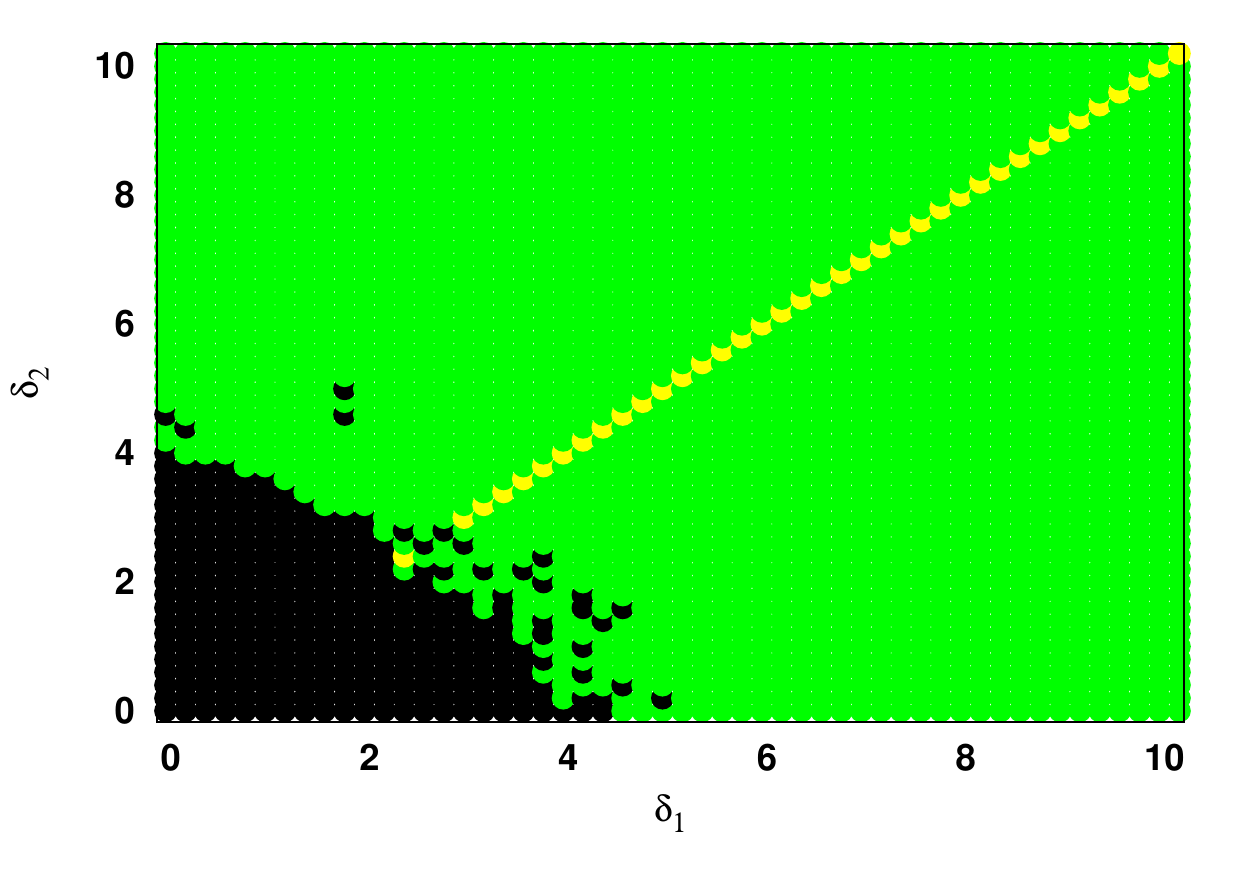} 
				\caption{Dynamical behaviour in the parameter space of delay  of first sub-region $\delta_{1}$ and the delay of the second sub-region $\delta_2$. Here the strengths of the self-delays of coupling is $\alpha_{1}=\alpha_{2} = 0.75$ for both sub-regions and the inter region coupling strength between both subsystem is $\gamma=0.1$ (for left figure) and $\gamma=0.4$ (for right figure). The black color represents amplitude death, red represents oscillator death, yellow represents homogeneous oscillations and green represents heterogeneous oscillations. Clearly, heterogeneous oscillations are predominant.}\label{phase_delay}
		\end{figure}

\begin{figure}[H]
				\centering
				\includegraphics[width=\twoFigureSize]{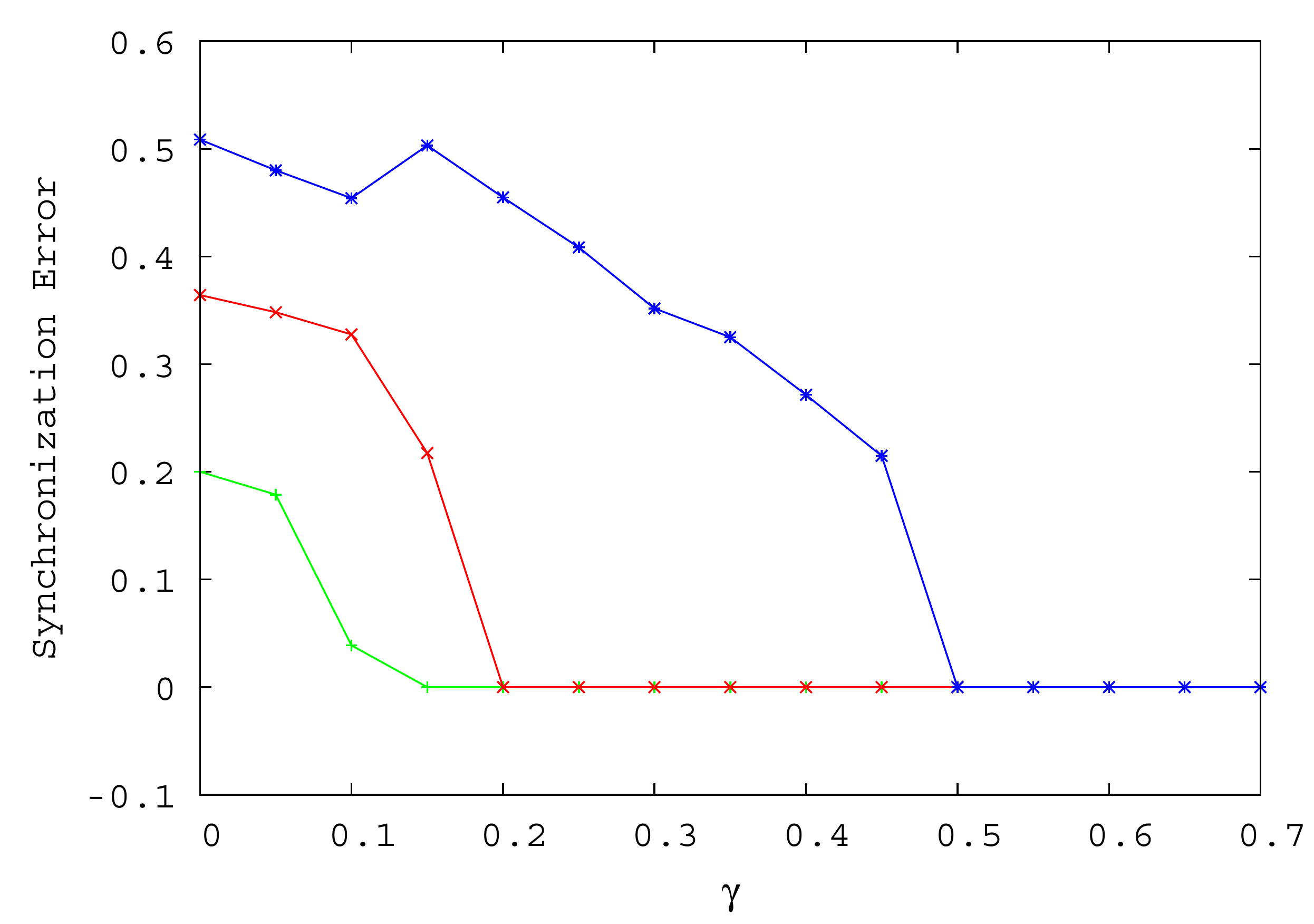}
				\caption{Synchronization error between the two sub-regions as a function of coupling strength $\gamma$, for $\delta_{2} = 1$ (green), $\delta_{2} = 2$ (red), $\delta_{2} = 5$ (blue). Here $\delta_{1} = 0$ and $\alpha_{1}=\alpha_{2} = 0.75$.}\label{synchronization_error}
		\end{figure}

		Interestingly, if we take the self-delay coupling strengths of the two sub-regions to be such that the temperature of one region goes to a fixed point regime when uncoupled, while the other system is in the oscillatory regime, then on coupling both systems show oscillations (see Fig. \ref{fp1}). This implies that oscillations may arise in certain sub-regions through coupling to neighbouring regions. For instance, it is clear from Fig. \ref{phase_delay} that a sub-region with very low delay ($\delta < 2$), which would naturally go to a steady state when uncoupled, yields oscillations (represented by the green color) when coupled to  another sub-region with high enough delay ($\delta > 2$).
		We also studied the synchronization of the oscillations in the two sub-regions with different delays, Fig. \ref{synchronization_error} shows representative results. We find that with increasing inter-region coupling strength $\gamma$, the synchronization error between $T_{1}$ and $T_{2}$ decreases, as expected. Further, when the differences in delay in the two regions is large, stronger inter-region coupling is necessary for synchronizing the sub-regions.		
		\begin{figure}[H]
				\centering 
				\includegraphics[width=\twoFigureSize]{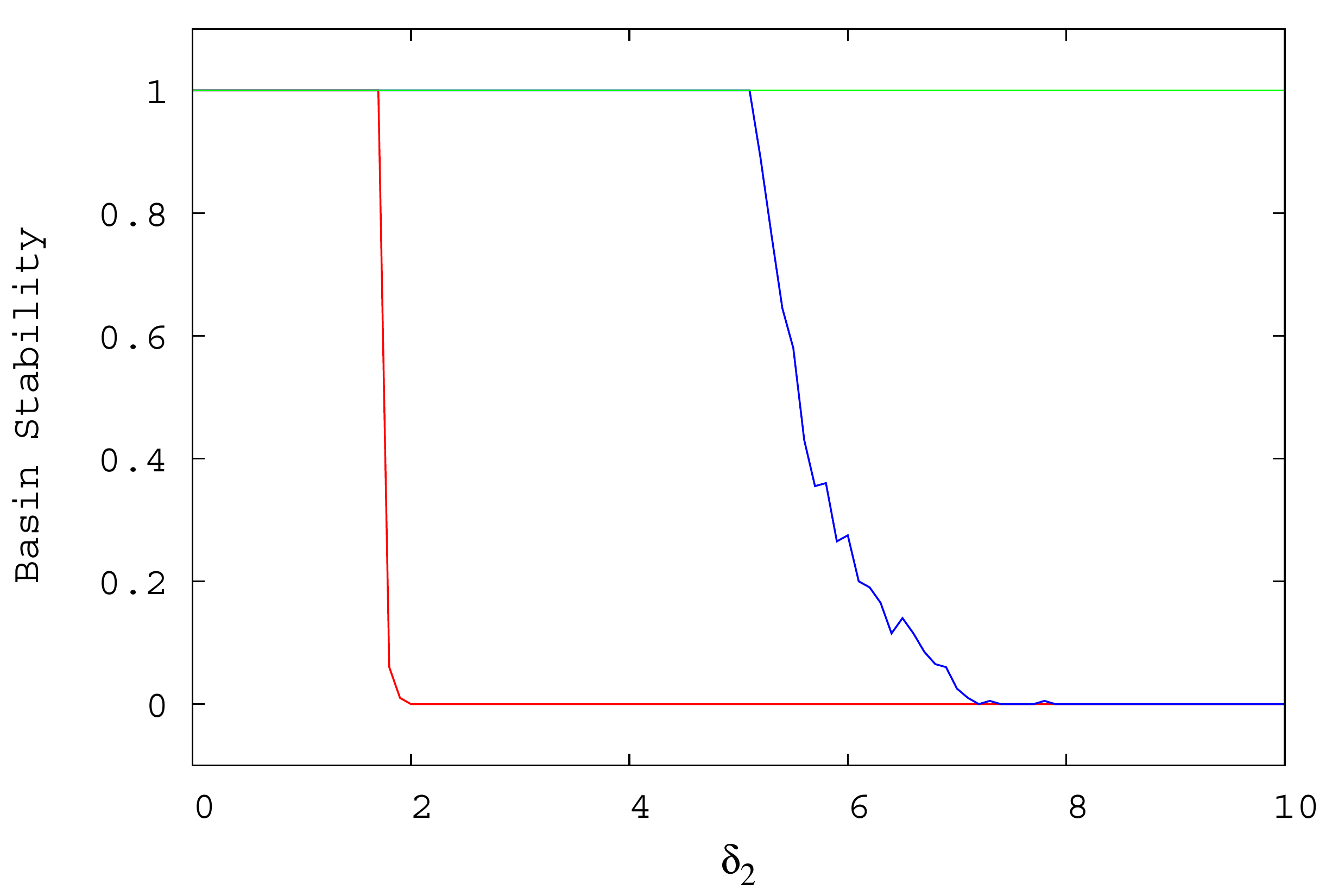} 
				\includegraphics[width=\twoFigureSize]{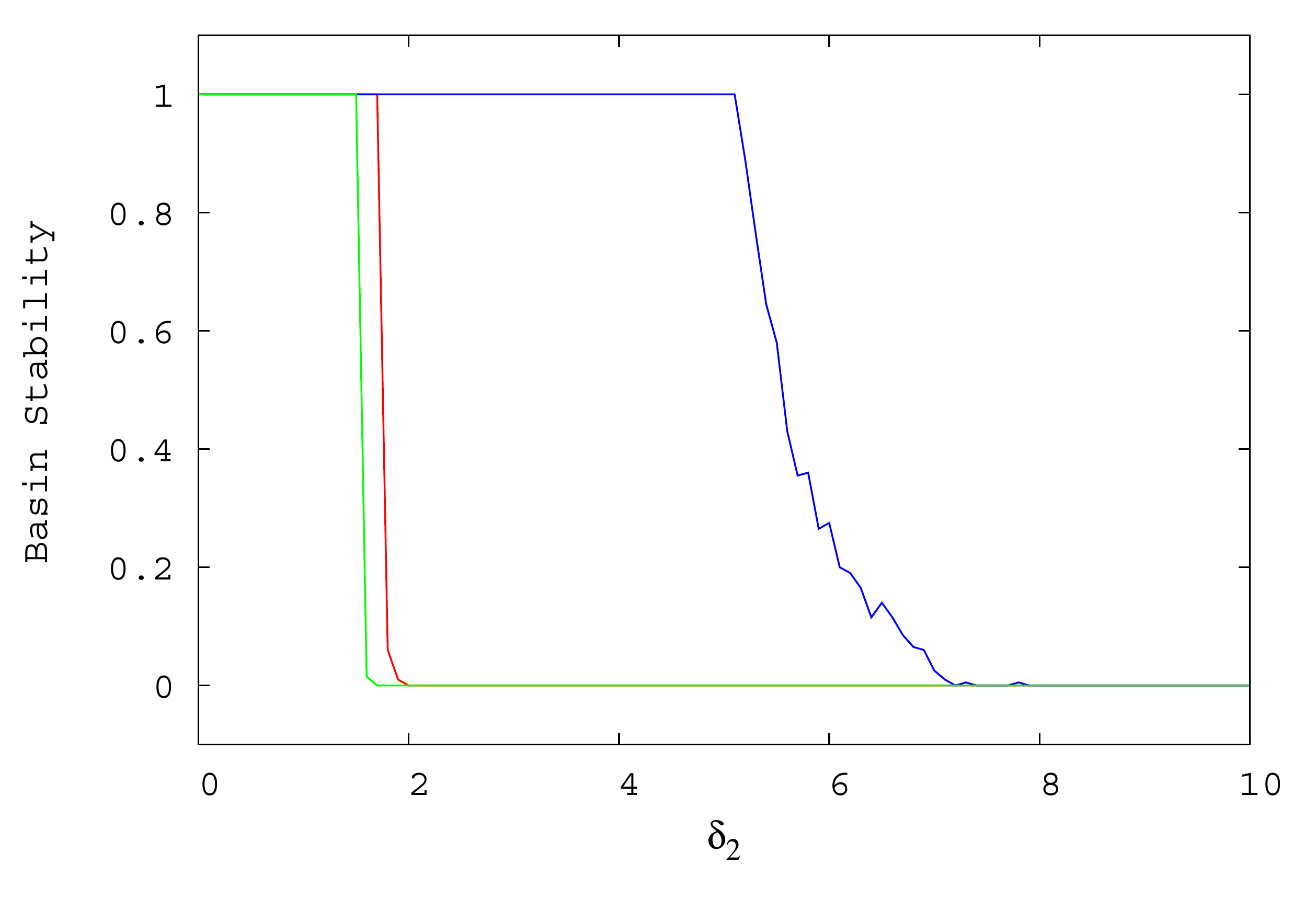}
	 
	(a)
	
				\includegraphics[width=\twoFigureSize]{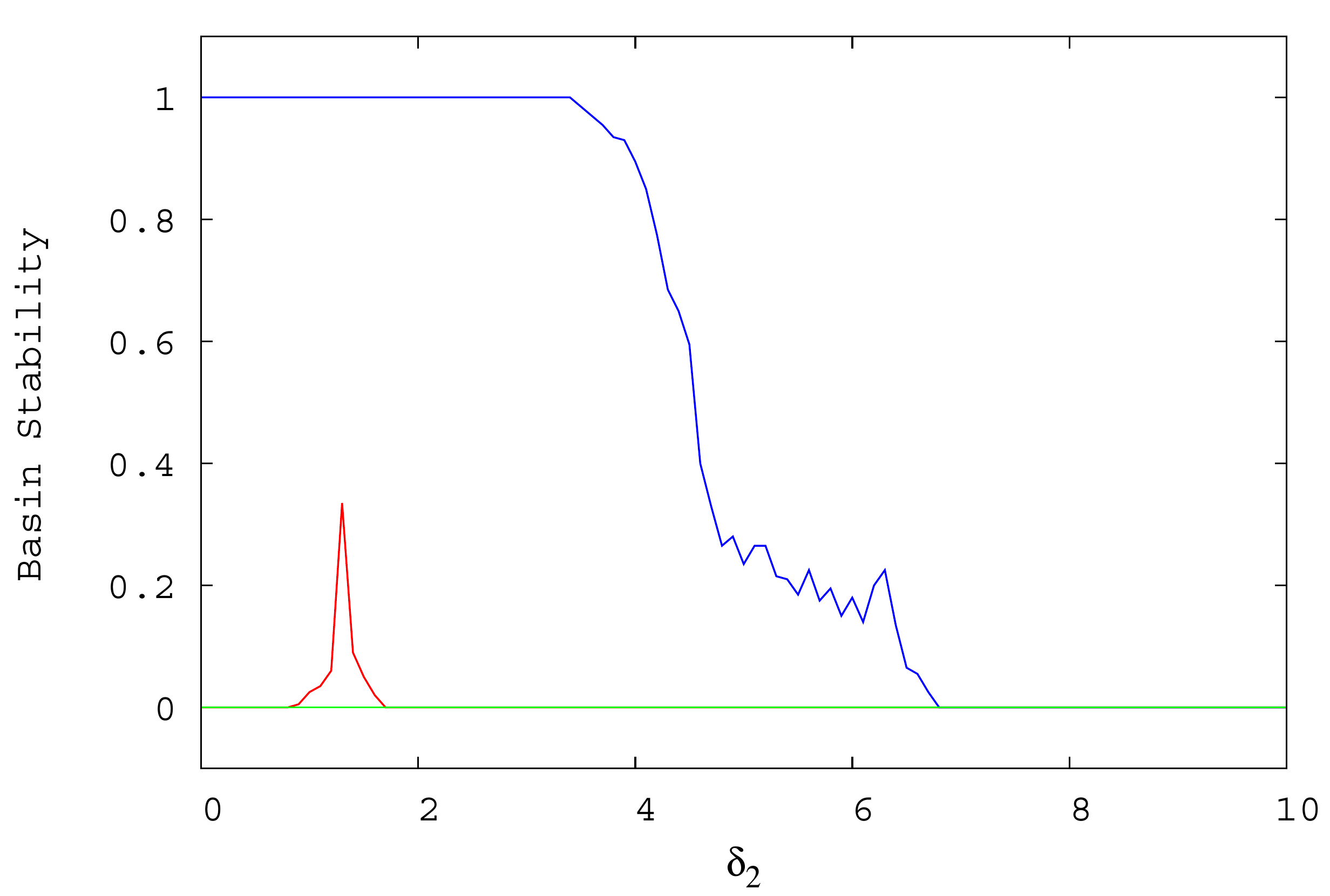} 
				\includegraphics[width=\twoFigureSize]{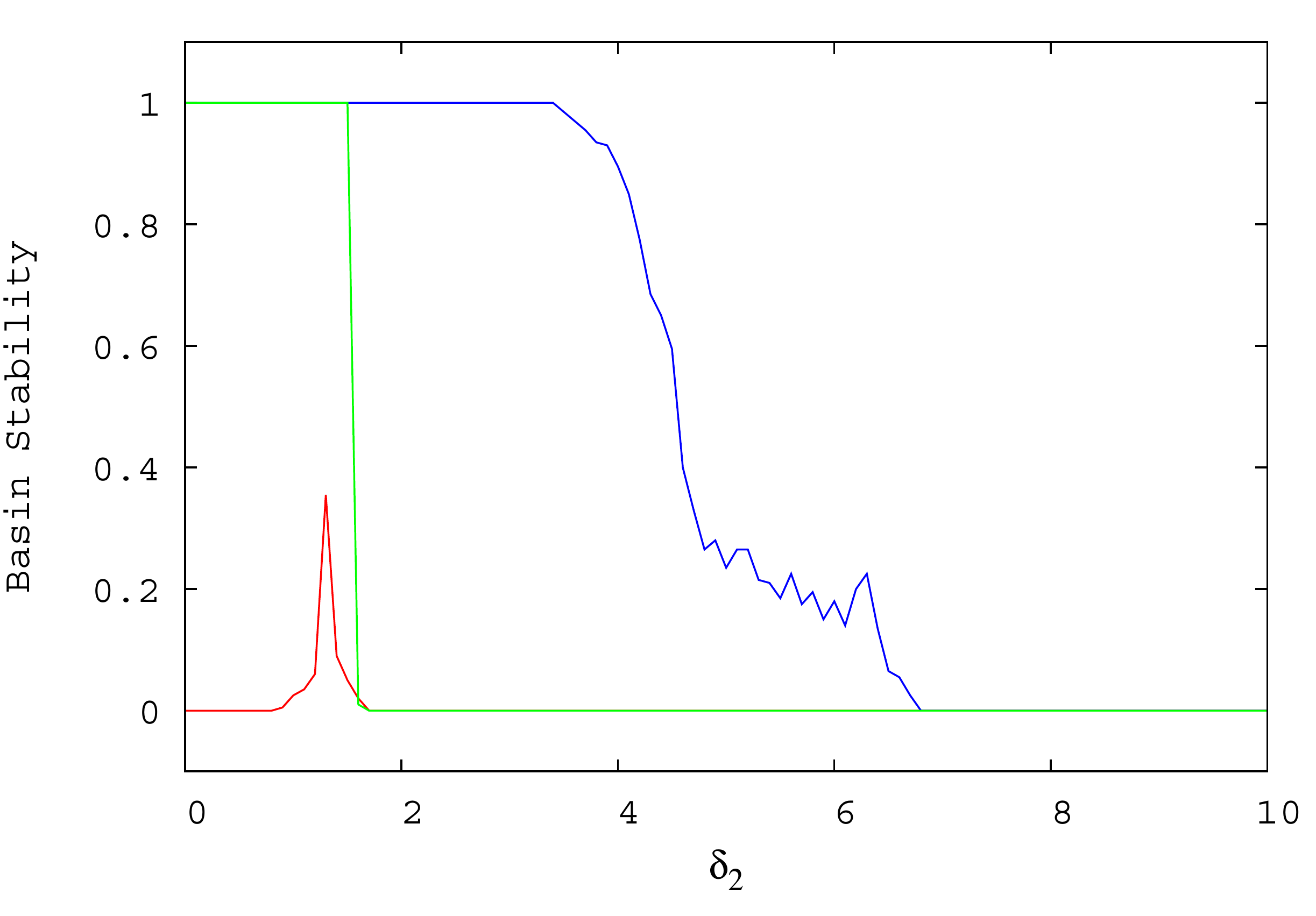} 
	
	(b)
				\caption{Basin stability for the fixed point as a function of $\delta_2$, for (left)  $T_{1}$ and (right) $T_{2}$. Here $\delta_{1}$ is (a) $0$ and (b) $2$. Three inter-region coupling strengths are displayed: $\gamma = 0$ (green), $\gamma =0.1$ (red) and $\gamma =0.5$ (blue). Here $\alpha_{1}=\alpha_{2} = 0.75$.}
				\label{basin}
			\end{figure}
				
	Lastly, the basin stability for the fixed point, as a function of $\delta_{2}$, keeping $\delta_1$ fixed, is displayed in Fig. \ref{basin}, for different values of inter-region coupling strengths. It is clear that high inter-region coupling $\gamma$ gives rise to co-existence of attractors. Also, as observed earlier, stronger coupling between sub-regions suppresses oscillations in larger regions of parameter space. 
					
	\section{Discussion}
	We have considered a system of coupled delayed action oscillators modelling the
	ENSO, and studied the dynamics of the sea surface temperature (SST) anomaly. 
	The existence and stability of the solutions arising in this model depend on three
	parameters: self delay, delay and inter-region coupling strengths. In our work we 
	explore the dynamics in the space of these parameters (cf. Table 1 for summary).
	The emergence or suppression of oscillations in our models is a dynamical feature of 
	utmost relevance, as it signals the presence or absence of ENSO-like behaviour.\\\\
	For identical sub-regions one typically observes a co-existence of amplitude and 
	oscillator death behavior for low delays, and heterogeneous oscillations for high 
	delays, when inter-region coupling is weak. For moderate inter-region coupling
	strengths one obtains homogeneous oscillations for sufficiently large delays and 
	amplitude death for small delays. When the inter-region coupling strength is large,
	oscillations are suppressed altogether, implying that strongly coupled sub-regions do
	not yield ENSO-like behaviour.\\
	Further we observe that larger strengths of self-delay coupling favours oscillations,
	 while oscillations die out when the delayed coupling is weak. This indicates again
	  that delayed feedback, incorporating oceanic wave transit effects, is the principal
	   cause of oscillatory behaviour. So the effect of trapped ocean waves propagating 
	   in a basin with closed boundaries is crucial for the emergence of ENSO-like behaviour.\\\\
	Note that in contrast to the well-known low order model of ENSO, the recharge
	oscillator and its important stochastic extensions, where the influence of the 
	neighbouring regions on the region of interest is modelled as external noise \cite{bianucci,recharge2}, 
	we consider neighbouring regions as a coupled deterministic dynamical systems.
	Different parameters yield a rich variety of dynamical patterns in our model,
	ranging from steady states and homogeneous oscillations to irregular oscillations,
	without explicit inclusion of noise.\\\\
	We also showed how non-uniformity in delays, and difference in the strengths of the self-delay coupling of the sub-regions, affect the rise of oscillations. The trends are similar to the uniform system. Namely, larger delays and self-delay coupling strengths lead to oscillations, while strong inter-region coupling kills oscillatory behaviour. The difference between the uniform case and the non-uniform system, is that amplitude death and homogeneous oscillations are predominant in the former, while oscillator death and heterogeneous oscillations are commonly found in the latter.
	Interestingly, we also find that when one sub-region has low delay and another has high delay, under weak coupling the oscillatory sub-region induces oscillations in sub-region that would have gone to a steady state if uncoupled.\\\\
	Further, in our system the effect of the strength of inter-region coupling depends on
	the specific features of the sub-systems. For instance, we observe that for 
	inter-region coupling strength, say $\gamma =0.15$ we have steady states when the 
	sub-systems are non-uniform, while stronger coupling, say $\gamma =0.25$, yields 
	oscillations for uniform sub-systems (cf. Table 1). Lastly, our dynamical model will
	 also help in providing a potential framework in which to understand synchronization
	 (or lack thereof) in the SST anomalies in different regions, which is an important
	 feature that has not yet been sufficiently explored.

	\begin{table}[H]
	
	\begin{tabular}{|p{2cm}|p{2cm}|p{2cm}|p{2cm}|p{3cm}|p{4cm}|} 
	\hline
	
	\multicolumn{6}{|c|}{ Table 1 }\\
	\hline
	$\alpha_{1}$ & $\alpha_{2}$ & $\delta_{1}$ & $\delta_{2}$ & $\gamma$ & Dynamics\\
	\hline
	$0.75$ & $0.75$ & $2$ & $2$ & $ \leq 0.25 $ & Oscillations \\
	$0.75$ & $0.75$ & $2$ & $2$ & $ >0.25 $ & Steady State \\
	$0.75$ & $0.75$ & $4 $ & $4$ & $ \leq 0.7 $ & Oscillations \\
	$0.75$ & $0.75$ & $4 $ & $4$ & $ >0.7 $ & Steady State \\
	\hline
	$0.75$ & $0.5$ & $ 2 $ & $2$ & $\leq 0.1$ & Oscillations \\
	$0.75$ & $0.5$ & $ 2 $ & $2$ & $ >0.1 $ & Steady State \\
	$0.75$ & $0.5$ & $ 4 $ & $4$ & $\leq 0.42$ & Oscillations \\
	$0.75$ & $0.5$ & $ 4 $ & $4$ & $ >0.42 $ & Steady State \\
	\hline
	$0.75$ & $0.25$ & $2 $ & $2$ & $\leq 0.1$ & Oscillations \\
	$0.75$ & $0.25$ & $2 $ & $2$ & $ >0.1 $ & Steady State \\
	$0.75$ & $0.25$ & $4 $ & $4$ & $\leq 0.27$ & Oscillations \\
	$0.75$ & $0.25$ & $4 $ & $4$ & $ >0.27 $ & Steady State \\
	\hline
	$0.75$ & $0.75$ & $1 $ & $2$ & $\leq 0.11$ & Oscillations \\
	$0.75$ & $0.75$ & $1 $ & $2$ & $ >0.11 $ & Steady State \\
	$0.75$ & $0.75$ & $1 $ & $4$ & $\leq 0.43$ & Oscillations \\
	$0.75$ & $0.75$ & $1 $ & $4$ & $ >0.43 $ & Steady State \\
	$0.75$ & $0.75$ & $2 $ & $4$ & $\leq 0.46$ & Oscillations \\
	$0.75$ & $0.75$ & $2 $ & $4$ & $ >0.46 $ & Steady State \\

	\hline
	
	\end{tabular}
	\caption{Summary of results from representative parameter values $\alpha_{1},\alpha_{2},\delta_{1},\delta_{2}$ and $\gamma$ (cf. Eqn. \ref{small_delay}).}
	\end{table}

\bigskip
	
	\noindent
	{\bf Acknowledgements}\\\\
	CM and SK would like to acknowledge the financial support from
	DST INSPIRE Fellowship, India.

	\end{document}